# Physics-Based Molecular Fingerprints from Spectral Graph Theory Provide Efficient Geometry-Aware Measures of Chemical Similarity

Jacob W. Toney[1,2], Ayleen Y. Farnood[1,2], Samir Darouich[1,3,4], and Heather J. Kulik[1,2,5,*]

[1]*Department of Chemical Engineering, Massachusetts Institute of Technology, Cambridge, MA 02139, USA*
[2]*Center for Computational Science and Engineering, Massachusetts Institute of Technology, Cambridge, MA 02139, USA*
[3]*Institute for Theoretical Chemistry, University of Stuttgart, 70569 Stuttgart, Germany*
[4]*Institute for Artificial Intelligence, University of Stuttgart, 70569 Stuttgart, Germany*
[5]*Department of Chemistry, Massachusetts Institute of Technology, Cambridge, MA 02139, USA*
*corresponding author email: hjkulik@mit.edu

ABSTRACT: Molecular representations are essential for the evaluation of molecular similarity and the development of structure–property relationships. Despite the known importance of three-dimensional (3D) structure to determine chemical and physical properties, the most widely used molecular fingerprints encode only two-dimensional connectivity. Such representations fail to distinguish similar but distinct stereoisomers and conformers. Alternative 3D methods are typically defined pairwise, making their application to large chemical spaces prohibitive, while deep learning embeddings are expressive but uninterpretable and limited by their training data diversity. Here, we introduce novel physics-inspired molecular fingerprints based on principles from spectral graph theory. We represent molecules as a complete graph in 3D space, with edge weights encoding heuristic physical interactions. Eigenvalue decomposition of the resulting graph Laplacian matrix results in a computationally efficient fixed-length chemical fingerprint that encodes 3D structure while obeying necessary physical symmetries of permutation and E(3) invariance. Spectral fingerprints differentiate between unique molecular structures with identical 2D connectivity, overcoming a limitation of 2D descriptors, while maintaining the low computational cost needed for efficient screening of vast chemical spaces. We evaluate our fingerprints with community detection algorithms and observe strong performance against representative baselines across datasets from organic, inorganic, biological, reticular, and reaction chemistry. Nearest-neighbor property estimation and applicability domain analyses reveal the utility of our molecular representation in machine learning and cheminformatics. We anticipate that spectral fingerprints will serve as generalizable, interpretable, and efficient measures of chemical similarity that incorporate 3D information at minimal cost.

## 1. Introduction

Modern computational chemistry relies on molecular representations. Early efforts focused on two-dimensional (2D) covalent bond structure, with SMILES strings developed as a simple, logical encoding of molecular structure.[1-3] The advent of machine learning (ML) has redefined the role of molecular fingerprints to serving as inputs for data-driven studies, with the chosen representations directly influencing the expressivity of parameterized models.[2-4] Widely used 2D representations such as Bemis–Murcko scaffolds[5] and extended connectivity fingerprints[6,7] are defined in terms of the molecular graph and substructures, explicitly encoding covalent bonding between atoms. While well-suited for small organic molecules, these methods generalize poorly to more diverse chemistry (e.g., organometallics, biological molecules, and macromolecules).[8-13] This has inspired the development of alternative connectivity-based representations such as the bag of bonds model[14], revised autocorrelations (RACs)[15,16], and various domain-specific encodings of coordination spheres and chirality[17-23]. Three-dimensional (3D) representations meanwhile incorporate spatial information, with widely employed examples including Coulomb matrices[24], atom-centered symmetry functions[25], and smooth overlap of atomic positions[26]. More recently, learned representations from deep learning models have emerged as alternatives to handcrafted descriptors. Although these task-specific embeddings achieve state-of-the-art performance on large datasets, their interpretability and generalizability are severely limited,[27-35] and their performance on small-to-moderate sized datasets remains inferior to descriptor-based methods.[31,36,37] Across this landscape, the choice of representation is critical, defining the space in which molecules are encoded and compared.

Central to cheminformatics and computer-aided molecular design is the similar property principle, which states that structurally similar molecules should exhibit similar properties.[38-40] Although no unique definition exists, chemical similarity is naturally defined in terms of a molecular representation and distance metric.[40-44] Each family of representations (e.g., 2D versus 3D, learned versus descriptor-based, etc.), however, exhibits a characteristic failure mode limiting their utility. 2D topological fingerprints are inexpensive but blind to geometry by construction, collapsing conformers and stereoisomers onto identical representations despite the known sensitivity of chemical properties to 3D structure.[8,9,45] Shape-based 3D similarity measures restore spatial awareness but typically require pairwise alignment of molecules.[46-49] This alignment requires a nonconvex optimization whose combinatorial analogue is NP-hard and which is ill-posed for molecules of differing elemental composition, prohibiting exhaustive pairwise comparison of all molecules in a large chemical space.[50-52] Alignment-free methods exist, but they rely on computationally expensive generation of conformer ensembles and explicit connectivity.[53] An ideal molecular representation will be inexpensive and interpretable while maintaining 3D awareness and generalizability across chemical domains. An effective fingerprint should further exhibit strong locality, embedding molecules such that proximity in representation space corresponds to similar properties.[14,54-57] Locality is the basis of cheminformatics and ML methods such as nearest-neighbor property estimation and uncertainty quantification, enabling the latter to serve as an effective measure for screening unlabeled datasets and estimating where a model is likely to fail.[57-63]

It is natural to represent molecules as graphs, where atoms and bonds correspond to vertices and edges, respectively.[64-68] Vertex attributes encode key atomic properties (e.g., atomic number), while edge weights typically reflect bond order.[2,4,69] One approach to incorporating 3D information in molecular graphs is to augment nodes with coordinate data, but this introduces sensitivity to a defined reference frame (i.e., translations and rotations). Alternative methods encoding pairwise atomic distances or interactions are sensitive to permutations of atomic indices (i.e., swapping atom indices alters the representation without changing the physics).[24,47,70] A useful molecular representation should be defined in 3D space, invariant to atomic permutation and to the choice of reference frame, avoid dependence on a bonding model, and retain favorable computational complexity. Spectral graph theory, i.e., the study of graphs in terms of their matrix representations and corresponding eigenvalues, offers a natural yet underexplored route to satisfying all these criteria simultaneously. The eigenvalues of a graph Laplacian are invariant to vertex relabeling by construction, are fixed in number once truncated, and encode both the global structure and the local connectivity of the underlying graph.[71-75] Earlier eigenvalue-based molecular representations include BCUT descriptors, which summarize selected eigenvalues of atom-property-weighted modified adjacency matrices, but they are derived from explicit 2D molecular connectivity.[76]

In this work, we introduce novel physics-based molecular fingerprints derived from spectral graph theory. We represent molecules as weighted complete graphs over 3D atomic coordinates, with edges weighted according to heuristic, physics-inspired interactions defined from tabulated atomic properties. Diagonalizing the resulting channel-wise (i.e., for each property evaluated) graph Laplacians and retaining the extremal eigenvalues yields a fixed-length fingerprint that is permutation and E(3) invariant and defined in 3D space, at a cost negligible relative to alignment-based or electronic-structure-derived alternatives. We validate our fingerprints on chemical systems representative of conformers, stereochemistry, reactions, and macromolecules, including relevant structure pairs indistinguishable by 2D topological methods. We integrate our fingerprints with community detection algorithms for efficient partitioning of large chemical spaces, benchmarking against 2D and 3D fingerprints across six datasets spanning organic, inorganic, biological, reaction, and reticular chemistry. Finally, we demonstrate the application of our spectral fingerprints to cheminformatics and ML workflows through *k*-nearest-neighbor regression and applicability-domain analysis, establishing them as a training-free complement to ML models.

## 2. Computational Methods

### Data Preparation

The spectral fingerprint was initially developed based on tests on several chemical systems including geometric conformers (i.e., boat vs. chair cyclohexane), stereochemistry (e.g., *cis* vs. *trans* symmetries), macromolecules (e.g., proteins), and reactions (see Sec. 3b). 3D structures of all representative chemical systems considered were prepared manually in Avogadro[77,78] version 1.2.0 and optimized with the Universal force field (UFF).[79] Each structure was then further optimized with density functional theory (DFT) at the ωB97M-V/def2-TZVP level of theory[80,81] using ORCA 6.1.0.[82-84] To analyze large chemical spaces, representative datasets from organic (QM9[85]), inorganic (tmQMg[86]), biological (QCell[87]), reaction (Transition1x[88]), and reticular (QMOF[89,90]) chemistry were selected from the literature (see Sec. 3c, Supporting Information Text S1). The QM9 dataset contains 133,885 3D structures representing small molecule organic chemistry.[85] While multiple versions of tmQMg exist with various sizes, we used the original version of the dataset with 60,799 structures as described in the corresponding publication (i.e., before addition of new complexes and removal of erroneous structures).[86,91] The lipids subset of the QCell dataset includes 16,000 fatty acid monomers, dimers, and trimers.[87] The Transition1x dataset includes reactant, product, and transition state geometries for 10,073 reactions.[88] The QMOF dataset contains results of DFT calculations on 20,372 experimentally-synthesized metal-organic frameworks.[89,90] To obtain xyz files, each of the cif files in QMOF was read using molSimplify version 1.8.0.[92,93] As a scaling test of spectral fingerprints to large datasets, we additionally consider the GEMS dataset of 2,713,986 off-equilibrium protein fragments and electronic properties.[94]

Spectral fingerprints were successfully generated for all entries of the datasets considered. Standard revised autocorrelations (RACs) and Coulomb-decay revised autocorrelations (CD-RACs)[15,16] were constructed with molSimplify[92,93] version 1.8.0 (Supporting Information Text S2). These descriptors exhibited the same parsability rates as our spectral fingerprints (i.e., 100% success for all datasets). Each structure was also processed using RDKit[95] version 2026.3.1 to construct Morgan fingerprints[6,7] and Bemis–Murcko scaffolds.[5] Structures unparsable by RDKit were omitted from subsequent analyses. Each dataset considered was clustered separately using spectral, RACs, CD-RACs, and Morgan representations. A similarity graph was constructed with nodes corresponding to fingerprints and each node connected to its $k = 5 \log N$ nearest neighbors. Edges were weighted by the pairwise similarity between nodes, calculated using inverse Euclidean distance for continuous fingerprints (i.e., spectral, RACs, and CD-RACs) or Tanimoto similarity for binary fingerprints (i.e., Morgan). The resulting similarity graphs were segmented into distinct clusters with the Leiden community detection algorithm, optimizing for modularity according to Reichardt and Bornholdt's Potts model[96,97] with a resolution of 1 as implemented in leidenalg 0.11.0.[98] As no numerical representation is provided from scaffold-based splits, the molecular scaffolds themselves are considered to represent clusters. All analysis scripts are available in the corresponding Zenodo repository.[99]

For each dataset and representation, we evaluate cluster quality in latent (i.e., fingerprint) and target (i.e., property of interest) space. Specifically, we calculate the percentage of cluster-pairs with statistically significant differences in either latent space or target distributions according to Mann–Whitney U test[100] statistics at the $\alpha = 0.05$ significance level as implemented in SciPy[101] 1.15.2. We also report Davies–Bouldin indices[102] to measure intra-cluster compactness and inter-cluster separation, as well as Calinski–Harabasz indices[103] indicating the ratio of inter- to intra-cluster variance. Silhouette scores[104] are also calculated, though are expected to be less meaningful for non-convex clusters and community detection (Supporting Information Table S1).[105,106] Davies–Bouldin indices, Calinski–Harabasz indices, and Silhouette scores are calculated as implemented in scikit-learn[107] 1.7.2.

**Machine Learning Models**

All machine learning models reported in this work were trained using the ElemeNet software package.[108] For QM9, training, validation, and test splits from Heid et al. were used.[109] For tmQMg, partitions from Kneiding et al. were used, including by excluding the same 2,390 erroneous structures identified and omitted by the authors.[86] As in the original ElemeNet work, models were trained on QM9 atomization energy, dipole magnitude, HOMO-LUMO gap, and polarizability, in addition to tmQMg dipole magnitude, HOMO-LUMO gap, and polarizability. For each model, both 2D and E(3) equivariant 3D graph neural network (GNN) encoders were considered with either multilayer perceptron (MLP) or transformer readout architectures, resulting in a total of 4 models trained per property (i.e., two encoder architectures and two readout architectures). All models were trained using mean-squared error loss with the AdamW optimizer[110] for 1,000 (2D GNNs) or 2,000 (3D GNNs) epochs. Default hyperparameters were used for all models, with early-stopping based on validation loss (Supporting Information Table S2 and S3). 2D models use connectivity derived from either SMILES string (QM9) or mol2 (tmQMg) formats, while 3D models use xyz coordinates. Implicit hydrogens were used for all QM9 models. Property locality was quantified by leave-one-out (LOO) *k*-nearest-neighbor (*k*-NN) regression over the full QM9 and tmQMg datasets. For each molecule, the target property was predicted as the mean property of its *k*-nearest-neighbors (excluding self), with distance calculated in each representation space using the corresponding distance metric (i.e., Euclidean for continuous fingerprints, Tanimoto for binary fingerprints). Applicability-domain estimation was performed for each trained model by ranking each test set molecule by its distance to the nearest training set molecule, calculated independently in each representation space. Performance metrics are evaluated on test set molecules in order of increasing distance and reported as functions of distance for each model and representation space.

**Numerical Implementation of Spectral Fingerprints**

From a given molecular input file, adjacency matrices and graph Laplacians are constructed efficiently with NumPy[111] 2.3.3 and SciPy 1.16.2, leveraging sparsity for structures with over 1,000 atoms. Similarity is calculated pairwise and accelerated with just-in-time compilation and parallel processing available through Numba[112] 0.62.1 and joblib[113] 1.5.2. For large chemical databases (i.e., over 1,000 entries), our implementation is accelerated with the high-performance FAISS library[114,115] version 1.10.0 for similarity search across millions of vectors. Unsupervised clustering is performed with the Leiden algorithm[98] utilizing igraph[116,117] 1.0.0 and leidenalg 0.11.0. Supervised clustering with the *k*-means algorithm[118-120] as implemented in scikit-learn 1.7.2 is also supported. Clusters are visualized using cluster-specific projection of fingerprints by principal component analysis[121,122] in the manner outlined by Zsigmond and coworkers.[123] Atomic number, valence electrons, electronegativity, and covalent radius of each element was obtained from www.webelements.com[124] and https://periodictable.com.[125] The first ionization potential for each neutral element was extracted from Mendeleev[126] version 0.19.0, while the polarizability and van der Waals radius was taken from the primary literature.[127,128] Our codebase was developed and tested using Python 3.10.20.

## 3. Results and Discussion

### 3a. Mathematical and Physical Derivation

We first introduce several concepts from graph theory that will be used throughout this work. A graph $G$ is defined as a collection of vertices $V$ connected by edges $E$, denoted $G = \{V, E\}$. Any pair of vertices $(v_i, v_j)$ is considered connected if an edge $e_{ij}$ between them exists. In the most general case, edges are undirected ($e_{ij} = e_{ji}$) and unweighted. Simple graphs are those in which no self-loops exist (i.e., no vertex is connected to itself). A complete graph is one in which an edge exists for every unique pair of vertices, resulting in $N(N-1)/2$ edges for $N$ vertices. In this work, we exclusively consider undirected, simple, and complete graphs. We begin our discussion with unweighted edges before introducing edge weights.

For any simple graph $G$ with $N$ vertices, we may construct its corresponding adjacency matrix $A \in \mathbb{R}^{N \times N}$, where each element $a_{ij}$ indicates the presence of an edge between vertices $v_i$ and $v_j$. For an unweighted graph, $A$ is binary, with 1 indicating the presence of an edge. The adjacency matrix of a simple graph always has all zero diagonal elements ($a_{ii} = 0$). While the adjacency matrix encodes pairwise connectivity, the degree matrix $D$ encodes the total connectivity of each vertex (eqn. 1). The degree of a vertex is defined as the number of vertices with which it shares an edge, so $D$ is diagonal with entries:

$$d_{ii} = \sum_{j=1}^{N} a_{ij} \tag{1}$$

A key object in graph theory is the graph Laplacian (eqn. 2), defined as:

$$L = D - A \tag{2}$$

The Laplacian of a simple, undirected graph is positive semidefinite, so all eigenvalues are nonnegative. The eigenvalues of $L$ are referred to as the spectrum of the graph, encoding both global structure and local connectivity.[71]

We extend our unweighted molecular graph to a weighted complete graph, where instead of encoding only covalent bond order, we assign edge weights based on pairwise atomic interactions (Figure 1). This is motivated by the need to capture the full three-dimensional chemical environment. Each type of distance-dependent atomic interaction is assigned a separate channel, and their nature will be described in more detail next. Bond-connectivity graphs discard long-range noncovalent interactions that are critical to properties such as reactivity, solubility, and molecular recognition. Because each channel weight depends on interatomic distance $r_{ij}$, the representation encodes 3D geometry beyond connectivity. We introduce a multichannel adjacency tensor $\mathbf{A} \in \mathbb{R}^{N \times N \times 4}$ encoding four pairwise, physics-based interactions, each parameterized using atomic properties from a precalculated look-up table: electrostatics ($e$), bonding ($b$), sterics ($s$), and dispersion ($d$, eqns. 3 – 6). The electrostatics channel follows a Coulombic potential, measuring pairwise repulsion between total nuclear charges $Z_i$ and $Z_j$ separated by interatomic distance $r_{ij}$:

$$a_{ij}^{(e)} = \frac{Z_i Z_j}{r_{ij}} \tag{3}$$

The bonding channel is inspired by Slater-type orbitals and approximates overlap of valence electrons ($v$), weighted by differences in electronegativity ($\chi$)[129]:

$$a_{ij}^{(b)} = v_i v_j e^{-r_{ij}|\chi_i - \chi_j|} \tag{4}$$

The sterics channel is modeled via a soft-steric overlap function, taking a nonzero value only when the van der Waals radii $r_{\mathrm{vdW}}$ of atoms $i$ and $j$ overlap:

$$a_{ij}^{(s)} = \max\left(0, \frac{r_{\mathrm{vdW},i} + r_{\mathrm{vdW},j} - r_{ij}}{r_{\mathrm{vdW},i} + r_{\mathrm{vdW},j}}\right) \tag{5}$$

Finally, the dispersion interactions are encoded according to London's formulation using first ionization potentials ($IP$) and polarizabilities ($\alpha$)[130]:

$$a_{ij}^{(d)} = \frac{IP_i IP_j}{IP_i + IP_j} \frac{\alpha_i \alpha_j}{r_{ij}^6} \tag{6}$$

Each channel $A^{(k)}$ of this rank-3 tensor represents a standard adjacency matrix for the $k$-th physical interaction, denoted $A^{(e)}$, $A^{(b)}$, $A^{(s)}$, and $A^{(d)}$ (Figure 1). To enable comparison across molecules with different size and composition, pairwise atomic distances are normalized by the sum of covalent radii, while adjacency matrices are divided by their Frobenius norm to normalize their magnitudes.[131,132] It is important to note that these channels are heuristic, physics-inspired relationships designed to capture chemically relevant interactions at low computational cost.

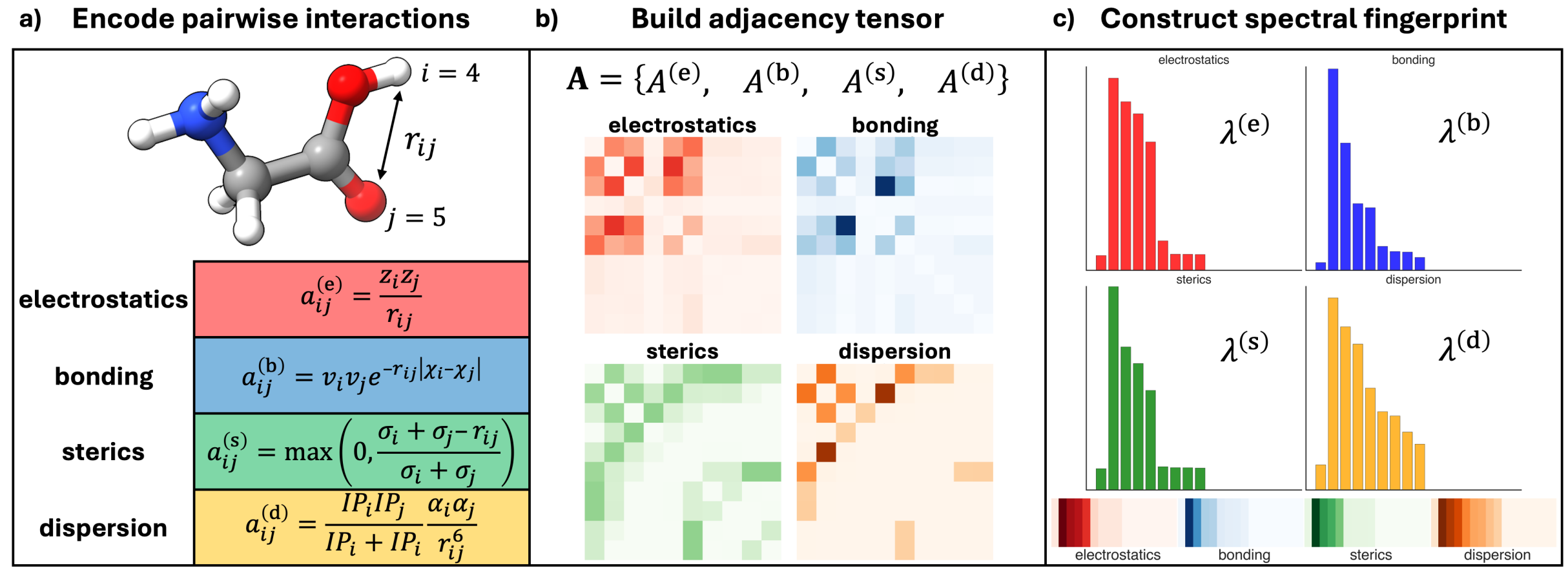

**Figure 1.** Illustrative procedure for generating spectral fingerprints from 3D atomic coordinates. a) 3D coordinates of glycine optimized with UFF. Functional forms for pairwise interactions between atom pairs $(i, j)$ are shown for electrostatic, bonding, steric, and dispersion channels. Atoms are colored as follows: H in white, C in gray, N in blue, O in red. b) Heatmaps of normalized adjacency matrices for electrostatic, bonding, steric, and dispersion interactions. Matrix dimensions are $10 \times 10$, corresponding to the 10 atoms in glycine. Matrix entries range from 0 to 0.5. c) Eigenvalues of graph Laplacians constructed from each adjacency matrix. We place the Fiedler eigenvalue (the smallest nonzero eigenvalue) first and sort all other eigenvalues in descending order starting from the largest. As glycine has only 10 atoms, each channel is padded with zeros to reach a length $M = 16$ eigenspectrum. The resulting values are concatenated into a length $4M = 64$ spectral fingerprint.

The multichannel Laplacian tensor **L** is constructed from **A** by channel-wise extension of standard Laplacian definitions (eqn. 7):

$$d_{ii}^{(k)} = \sum_{j=1}^{N} a_{ij}^{(k)}, \qquad L^{(k)} = D^{(k)} - A^{(k)}, \;\; \lambda^{(k)} = \mathrm{eig}\left(L^{(k)}\right), \qquad k \in \{e, b, s, d\} \tag{7}$$

For each channel $k$, the smallest Laplacian eigenvalue $\lambda_1^{(k)}$ is always zero.[71] If the corresponding positive-weight graph is connected, this zero eigenvalue has a multiplicity of one and the second-largest eigenvalue satisfies $\lambda_2^{(k)} > 0$. This second smallest value $\lambda_2^{(k)}$, known as the Fiedler eigenvalue or algebraic connectivity[72,73], encodes local connectivity, while the remaining eigenvalues capture the most significant chemical interactions. For molecules with $N \geq M$ atoms, we consider only the $M - 2$ largest eigenvalues (i.e., neglecting the trivially zero smallest eigenvalue $\lambda_1$ and the already included Fiedler eigenvalue $\lambda_2$). For molecules with $N < M$ atoms, we pad with zeros as necessary. This results in a length-$4M$ fingerprint $\gamma$ for each molecule (i.e., four physically motivated channels with $M$ eigenvalues each, eqn. 8):

$$\gamma^{(k)} = \left[\lambda_2^{(k)} \quad \lambda_N^{(k)} \quad \lambda_{N-1}^{(k)} \quad \cdots \quad \lambda_{N-(M-2)}^{(k)}\right], \qquad \gamma = [\gamma^{(e)} \quad \gamma^{(b)} \quad \gamma^{(s)} \quad \gamma^{(d)}] \tag{8}$$

We choose $M = 16$ as a default, balancing a small fingerprint size with chemical resolution allowing for comparison across molecules with different numbers of atoms. This value was determined empirically by considering a range of values from 2 to 128, though larger chemical systems beyond the scope of small molecules would benefit from dataset-specific tuning to determine the optimal value of $M$ (Supporting Information Figures S1 and S2). Similarity between molecules $i$ and $j$ is defined using the Euclidean distance between their corresponding spectral fingerprints (eqn. 9):

$$S_{ij} = \frac{1}{1 + \left\|\gamma_i - \gamma_j\right\|_2} \tag{9}$$

In addition to being grounded in interpretable physical interactions, our spectral fingerprints also exhibit several mathematical properties desired in a representation and similarity measure. The fingerprints are continuous, invariant to atomic permutations, and fixed-length regardless of molecular size. Though prior work has investigated Coulomb matrix eigenvalues for encoding molecular shape[133], our electrostatics channel follows a different functional form, while our bonding, sterics, and dispersion channels encode further interactions valuable in diverse chemical environments (Supporting Information Figure S3). The alignment-free nature of our spectral fingerprints also allows screening of large chemical databases with reduced time and memory requirements. Their definition in 3D space captures stereochemical information critical to chemical properties, as we demonstrate next.

### 3b. Demonstration on Chemically Relevant Examples

To illustrate the ability of spectral fingerprints to detect molecular similarity through sensitivity to three-dimensional structure, we first consider geometric conformers. A canonical example from organic chemistry is cyclohexane, which exhibits a chair conformer that is thermodynamically and kinetically favored over its twisted boat conformer due to reduced torsional and steric strain (Supporting Information Table S4).[134] We prepared and geometry optimized with DFT (see Sec. 2) 3D structures of cyclohexane chair and twisted boat conformers, and then we determined their spectral fingerprints. The pairwise similarity between spectral fingerprints of the chair and twisted boat conformers is 0.961 (i.e., they are distinguishable). For reference, the pairwise similarity between chair conformers of cyclohexane and piperidine is 0.678, consistent with higher chemical similarity between conformers than for element substitutions. Beyond conformational isomers, we also examine the square planar and tetrahedral coordination geometries commonly observed in transition metal chemistry (Supporting Information Table S4). Due to the diversity of metal–ligand combinations, oxidation states, and electron configurations, coordination complexes may adopt multiple VSEPR geometries (e.g., octahedral, trigonal bipyramidal, etc.) despite identical molecular graphs.[135] We applied an analogous procedure to bis-chelate Ni(II) complexes known to undergo square planar–tetrahedral isomerization upon changing between singlet and triplet spin states.[136] The square planar and tetrahedral Ni(II) complexes exhibit a similarity of 0.812 (Figure 2). These results demonstrate that spectral fingerprints correctly identify these molecular pairs as highly similar yet structurally distinct using only their 3D coordinates. Notably, each pair is indistinguishable from a 2D topological perspective, causing commonly used 2D fingerprints such as Bemis–Murcko scaffolds[5] and Morgan fingerprints[6,7] to assign identical representations to members of each pair and underscoring the necessity of 3D-aware molecular representations.

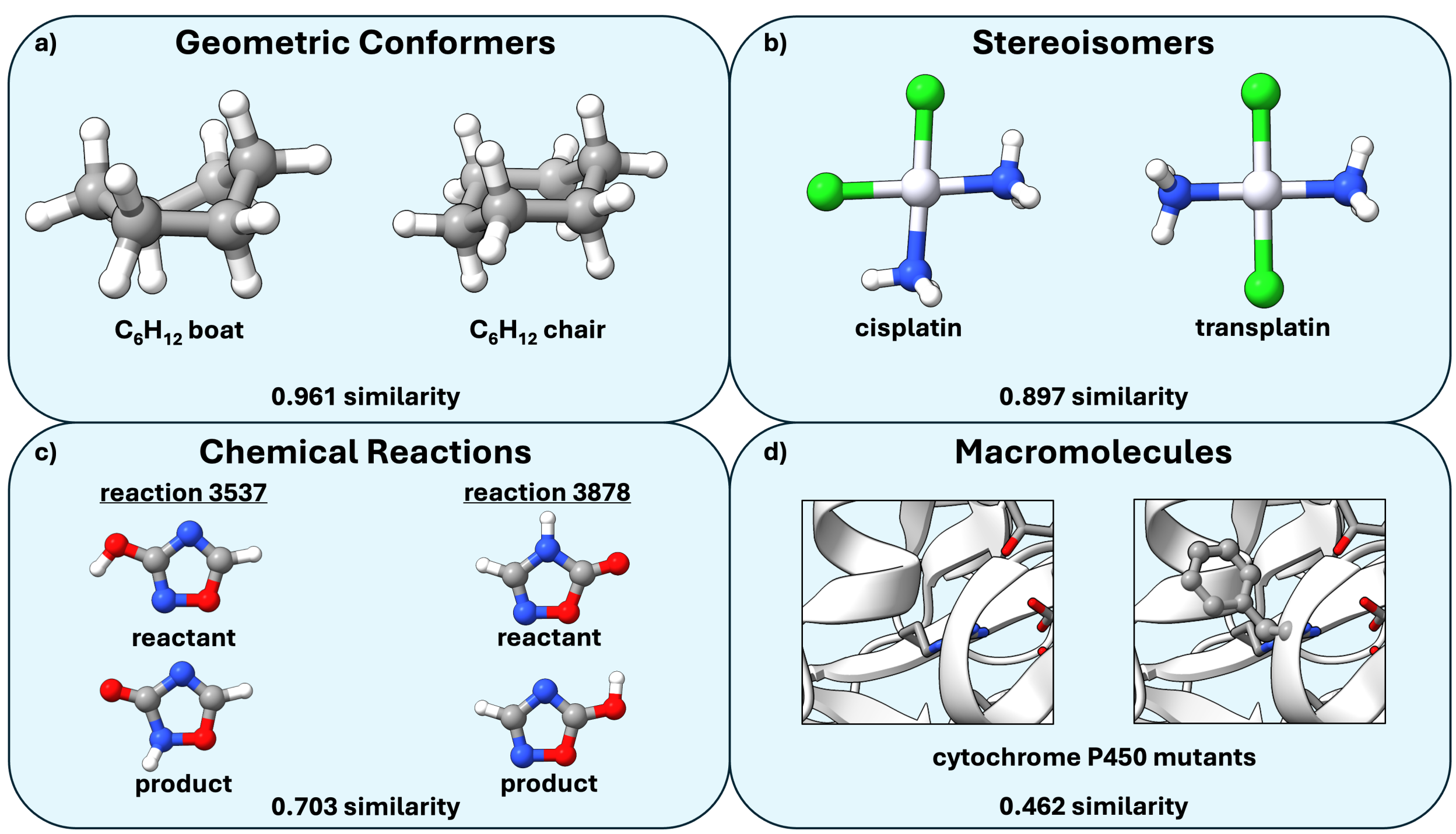

**Figure 2.** Representative chemical systems analyzed using spectral fingerprints. a) Boat and chair conformers of cyclohexane ($C_6H_{12}$) exhibit a high spectral similarity of 0.961. b) *cis* and *trans* stereoisomers of Pt(II)$Cl_2(NH_3)_2$ are described by distinct spectral fingerprints. c) Reaction spectral fingerprints between (reactant, product) pairs of oxidation and reduction reactions from the Transition1x dataset. d) *in silico* Gly-to-Phe mutation performed on cytochrome P450 (PDB ID: 8U1I) and resulting 0.436 spectral similarity. For the P450 enzyme example, only the Phe sidechain is shown explicitly; all other backbone atoms not shown. Atoms colored as follows: H in white, C in gray, N in blue, O in red, Cl in green, and Pt in silver. Protein backbone colored in light gray.

We also consider stereoisomers, where interconversion between configurations is prohibited by energetically unfavorable bond rotations. The classical example is *cis*/*trans* isomerism, in which substituents separated by a double bond occupy either the same (*cis*) or opposite (*trans*) sides, often resulting in substantial differences in molecular properties (Supporting Information Table S4). An example from organic chemistry is maleic acid and fumaric acid, each of which are derivatives of butenedioic acid.[134] While each molecule contains one carboxyl group on each of the two double-bonded carbon atoms, maleic acid adopts the *cis* configuration (i.e., both carboxyl groups on the same side), while fumaric acid exhibits carboxyl groups in the *trans* position. This leads to dramatically different molecular properties: maleic acid is readily water-soluble and used in polymer resins, while fumaric acid is relatively insoluble and used as a food additive.[137,138] The spectral similarity between these two isomers is 0.869, while 2D fingerprints yield identical representations and a similarity of 1. An additional relevant example is offered by cisplatin and transplatin. These coordination complexes exhibit the same molecular formula, $Pt(II)Cl_2(NH_3)_2$, and identical connectivity. However, while cisplatin is a potent chemotherapy medication, transplatin is considered clinically inactive.[139-141] Cisplatin and transplatin are observed to have a 0.897 spectral similarity score. Finally, we examine the case of facial (*fac*) and meridional (*mer*) isomers, distinct symmetries accessible to octahedral transition metal complexes.[135] As a representative example, we consider homoleptic iridium complexes with three bidentate 2-phenylpyridine (ppy) ligands (Supporting Information Table S4).[142] The spectral similarities computed for the *fac*/*mer* $Ir(III)(ppy)_3$ isomer pair is 0.956 (Figure 2). These examples further illustrate that spectral fingerprints consistently identify stereoisomers as closely related, yet distinct.

While the preceding examples validate the ability of spectral fingerprints to distinguish between closely related structures, an effective molecular representation must also capture broader chemical trends and generalize to more complex chemistries. We therefore extend our analysis to progressively more challenging cases, including chemical reactions, extended materials, and proteins. Because a complete graph can be constructed for any collection of atoms defined in 3D space, our framework extends naturally to reactions involving multiple reactant and product species. The permutation invariance of our spectral methods is particularly advantageous as it eliminates the need for explicit atom mapping between reactants and products. Reaction fingerprints are constructed by concatenating the individual length-$M$ spectral fingerprints of the reactants and products, yielding a fixed-length vector of dimension $8M$. Pairwise reaction similarity is then be computed analogously to molecular similarity, using the Euclidean distance between spectral fingerprints of each reaction. Provided that the relative order of species is maintained, any combination of $N$ molecular species may be represented in a length $4MN$ spectral fingerprint. This formulation naturally generalizes to encoding chemically relevant species, such as solvents, catalysts, intermediates, or transition states.

As a demonstration, we examine three reactions from the Transition1x dataset of organic reactions modeled at the hybrid DFT level of theory.[88] Specifically, we first consider reactions 3537, corresponding to the oxidation of 1,2,4-oxadiazol-3-ol to 1,2,4-oxadiazolone, and 3878, a ketone reduction reaction between 5-oxo-1,2,4-oxadiazoline and 2,4-dihydro-1,2,4-oxadiazol-4-ium-2-id-5-ol (Supporting Information Table S4). The reported activation energies for these reactions are both approximately 2.6 kcal/mol. We additionally consider reaction 0728, a ring-opening transformation of pyrazol-3-ylamine to an acyclic species, which has a higher activation energy of 6.4 kcal/mol (Supporting Information Table S4). We compute spectral fingerprints for each reaction by concatenating the reactant and product fingerprints, giving a similarity score of 0.703 for the two alcohol-to-ketone transformations (Figure 2). Meanwhile, the similarities between each oxidation/reduction reaction and the ring-opening reaction are substantially lower at 0.434 and 0.429. This similarity measure mirrors the difference in activation energies $\Delta\Delta E^{\ddagger}$ between each reaction. The two oxidation and reduction reactions differ by only $\Delta\Delta E^{\ddagger} = 7\text{E-}3$ kcal/mol, whereas ring-opening reaction differs from each by $\Delta\Delta E^{\ddagger} = 3.9$ kcal/mol. This suggests that spectral fingerprints can be used to distinguish reactions without any use of the transition state geometry, underscoring the utility of these methods in screening chemical databases of both molecular and reaction properties.

We next demonstrate that our spectral fingerprints scale to systems larger than the molecules considered above, specifically, to larger, periodic materials systems such as metal-organic frameworks (MOFs) containing hundreds of atoms per unit cell. Towards this end, we consider entries from the QMOF dataset[89,90], a collection of MOFs curated from the Cambridge Structural Database[143] (CSD) with electronic properties calculated by periodic DFT calculations. We select two zinc-based MOFs with reasonably similar structures: one with imidazole and phthalate-based linkers (CSD refcode ACOLIP, QMOF ID 355e9d2) and one with 2,6-dimethyl-4-styrylpyridine-3,5-dicarboxylate linkers (CSD refcode HAXMUQ, QMOF ID 21b3989). We further consider a gadolinium MOF with acetate linkers (CSD refcode HAVFAN, QMOF ID 4c0ef7e), exhibiting more distinct metal and linker chemistry (Supporting Information Table S4). While these unit cells each contain over 100 atoms on average, our spectral fingerprints effectively scale to such extended systems. We observe that the Zn MOFs are ranked with a similarity of 0.740, consistent with their similar band gaps of 2.7 and 2.8 eV, respectively. In contrast, the lanthanide MOF is ranked as dissimilar from both the Zn systems (similarity between 0.267–0.277) in agreement with its distinct chemistry and significantly larger band gap of 4.9 eV.

As a final demonstration of our spectral fingerprints on complex macromolecular systems, we consider two pairs of proteins extracted from the Protein Data Bank[144] (PDB). Due to their large size, we expected these systems to serve as a challenging benchmark for our spectral fingerprints from both a chemical and algorithmic perspective. Conventional methods for calculating protein similarity rely on sequence or atom alignment, which scale poorly with both protein size and the number of proteins being screened.[145,146] While spatial and sequence alignment-free alternatives exist, these typically require consistent ordering of atom indices[147,148] or are string- or graph-based representations which do not explicitly encode 3D chemical information.[149-151] Representing a system of thousands of atoms with a length-64 spectral fingerprint constitutes a dramatic dimensionality reduction, and it is unclear *a priori* whether our heuristic descriptions of electrostatics, bonding, sterics, and dispersion will encode relevant phenomena across such size and length scales. Furthermore, the eigenvalue decomposition required to generate our fingerprints scales as $O(N^3)$ for an $N$-atom system, becoming the computational bottleneck. To make matrix construction tractable for macromolecules, we use a 10 Å cutoff when computing pairwise atomic interactions and represent the resulting distance matrix with the compressed sparse row matrix format for systems with over 1,000 atoms.[152]

To test our method, we chose a cytochrome P450 protein (PDB ID: 8U1I), a heme-containing enzyme with over 3,600 atoms. We performed *in silico* mutations designed to result in varying degrees of structural distortion to the protein (Supporting Information Table S4). Specifically, we replaced all 31 glycine (Gly) residues with alanine (Ala), which differ by only a single methyl group (i.e., a Gly-to-Ala mutation), resulting in an increase in Ala residues from 45 to 76 in the mutant protein. We also separately replaced all Gly with phenylalanine (Phe, selecting the most prevalent Dunbrack rotamer in all cases[153]), which we expected to more greatly distort the protein from its wild-type structure (i.e., Gly-to-Phe mutation), resulting in an increase from 15 to 46 total Phe residues in the mutant protein (Supporting Information Table S5). Despite their including ten times as many atoms as the largest systems studied thus far, spectral fingerprints are successfully generated for all three of these systems. For the Gly-to-Ala mutation, we observe a similarity of 0.976 to the wild-type P450 protein. This is consistent with their differing by only a single methyl group at each of the 31 mutation sites with each methyl group projected along the same bond vector as the hydrogen atom it replaces in the wild-type protein structure. This is in contrast to the Gly-to-Phe mutation, which exhibits a similarity to the wild-type of only 0.462 because the phenyl residues now orient in a distinct fashion throughout the protein (Figure 2). This is indicative of the fact that spectral fingerprints distinguish between proteins with a fixed backbone that differ only in a subset of amino acid side chains. We repeated this analysis for the non-heme iron enzyme WelO5 (PDB ID: 5J4R) and observed similar results (i.e., 0.971 similarity to wild-type for Gly-to-Ala mutation, 0.436 similarity to wild-type for the Gly-to-Phe mutation, Supporting Information Table S4). Although fingerprint generation is more computationally expensive for these large systems than for small molecules, the process scales effectively to these thousand-atom systems (Supporting Information Table S6). Together with our observations on geometric conformers, stereoisomers, and chemical reactions, these results indicate that spectral fingerprints remain discriminative across diverse chemical systems. We now turn to the application of these fingerprints for screening large chemical datasets.

### 3c. Clustering Algorithms for Screening Large Chemical Databases

So far, we have demonstrated the ability of spectral fingerprints to encode molecular similarity. However, similarity is inherently a pairwise measure, with generating a full $N \times N$ similarity matrix exhibiting unfavorable $O(N^2)$ scaling with dataset size. In contrast, computing spectral fingerprints scales linearly with number of molecules, making the representation tractable for millions of molecules. To extend our analysis to such large chemical datasets, we leverage clustering algorithms that avoid explicit computation of all pairwise similarities. In particular, we employ the Leiden algorithm, a community detection method widely used in social networks and bioinformatics for partitioning a network into clusters of self-similar vertices.[98,154-159] Our implementation of the Leiden algorithm operates on a similarity graph where vertices represent spectral fingerprints and edges encode pairwise similarity. Each vertex is connected to its $k$ nearest neighbors, where $k = 5 \log N$, giving our approach for generating clusters without explicit pairwise screening of all members favorable $O(N \log N)$ scaling for a system with $N$ molecules. Together, these advantages enable our methods to efficiently screen large chemical databases, detecting molecular similarity entirely on the basis of spectral fingerprints. We now turn towards diverse datasets across chemical domains to benchmark the performance of spectral fingerprints against representative molecular representations.

We first evaluate our spectral fingerprint-based clustering algorithms on relevant datasets from organic, inorganic, and biological chemistry: QM9, tmQMg, and QCell. QM9 is a collection of 133,885 small organic molecules including up to 9 heavy atoms with only H, C, N, O, and F atom types.[85] tmQMg includes 60,799 mononuclear transition metal complexes (TMCs) from the Cambridge Structural Database (CSD).[86] Finally, the QCell dataset includes molecular dynamics trajectories of small biological molecules and fragments, including lipids, carbohydrates, and nucleic acids (Supporting Information Text S1).[87] For each dataset, we perform clustering in terms of molecular fingerprints. We also select a property of interest from each dataset (e.g., atomization energy, frontier orbital gap, dipole moment magnitude, etc.) and evaluate cluster quality using both extrinsic (i.e., global) and intrinsic (i.e., local) metrics (Supporting Information Text S3). Extrinsically, we report the percentage of cluster pairs with statistically distinct property distributions as determined by Mann–Whitney U tests. A higher percentage indicates that a greater proportion of the identified clusters correspond to chemically meaningful groups. We also report intrinsic metrics of the Davies–Bouldin index[102] (DBI), where lower values indicate more compact and well-separated clusters, and the Calinski–Harabasz index[103] (CHI), where higher values indicate better-defined cluster structure. Cluster quality benchmarks are performed in both latent (i.e., fingerprint) space and target (i.e., property of interest) space (Supporting Information Table S7). Clustering is only explicitly performed using chemical representations and thus is expected to exhibit stronger performance in latent space, while results in the more challenging target space are indicative of whether the identified clusters are chemically meaningful.

Generating spectral fingerprints across each dataset and clustering with our Leiden algorithm implementation results in clusters that are overall well-separated and reflective of distinct chemical properties (Table 1). Of the tasks, spectral fingerprints perform best on the QM9 atomization energy benchmark, with 98.0% of clusters identified to have distinct property distributions at the $\alpha = 0.05$ significance level. We also observe a low DBI of 156.6 and a high CHI of $9.65 \times 10^3$. Spectral fingerprints perform slightly worse on the tmQMg dataset relative to QM9, potentially due to the known presence of chemically erroneous outlier structures.[86,91] However, with 87.3% of clusters determined to have statistically distinct HOMO-LUMO gap distributions, the overall performance of our method is still strong. Spectral fingerprints result in clusters that correctly group two similar square planar Ni(II) complexes (CSD refcode BEPNOZ and SUXZUI) with identical first coordination spheres. These complexes have a pairwise spectral similarity of 0.934 and reside in the same cluster (Figure 3). The two complexes also have similar electronic properties, i.e., 0.01 hartree difference in HOMO-LUMO gap and 0.8 debye difference in dipole magnitude (Figure 3). Meanwhile, evidently dissimilar structures of a cobaltocene (CSD refcode KAGPAJ) and a linear gold halide (CSD refcode BESYOQ) with a spectral similarity of 0.190 are correctly assigned to distinct Leiden clusters.

**Table 1.** Key evaluation metrics for different clustering methods and datasets considered from organic (QM9), inorganic (tmQMg), and biological (QCell) chemistry. Extrinsic cluster quality is evaluated with respect to the percentage of cluster-pairs with statistically significant differences in their property distributions as determined by Mann–Whitney U tests.[100] Intrinsic quality is measured using the Davies–Bouldin index (DBI)[102] and Calinski–Harabasz index (CHI).[103] All metrics evaluated in target property space. Scaffold splits and Morgan fingerprints are generated using RDKit.[95] RACs[15] and CD-RACs[16] are generated using molSimplify.[92,93] Spectral fingerprints (our method) are observed to strike a balance between high performance and parsability. Bold indicates the best result for a given metric, and the arrow indicates what direction corresponds to a better value for a metric.

| Dataset (Domain) | Target (Units) | Representation | U-test (↑) | DBI (↓) | CHI (↑) | Parsable (%) | Clusters |
|---|---|---|---|---|---|---|---|
| QM9 (Organic) | Atomization Energy (kcal/mol) | Spectral | 98.0 | 157 | **9.65**×10$^3$ | 100 | 76 |
| | | Scaffold | 72.9 | 4.24×10$^3$ | 69.2 | 84.2 | 1132 |
| | | Morgan | 95.0 | 70.4 | 2.90×10$^3$ | 94.7 | 16 |
| | | RACs | **99.1** | 116 | 8.74×10$^3$ | 100 | 34 |
| | | CD-RACs | 98.2 | **36.7** | 9.37×10$^3$ | 100 | 34 |
| tmQMg (Inorganic) | Gap (eV) | Spectral | 87.3 | 586 | 145 | 100 | 31 |
| | | Scaffold | 73.6 | 762 | 31.0 | 70.9 | 230 |
| | | Morgan | **88.7** | **157** | **334** | 72.6 | 22 |
| | | RACs | 28.8 | 305 | 28.2 | 100 | 119 |
| | | CD-RACs | 61.1 | 278 | 103 | 100 | 68 |
| QCell (Biological) | Dipole Magnitude (D) | Spectral | 96.2 | 17.4 | **791** | 100 | 13 |
| | | Scaffold | **100** | 254 | 21.9 | 0.4 | 3 |
| | | Morgan | **100** | **2.39** | 188 | 20.0 | 3 |
| | | RACs | 88.1 | 241 | 391 | 100 | 35 |

| | | CD-RACs | 91.8 | 122 | 440 | 100 | 31 |
|---|---|---|---|---|---|---|---|

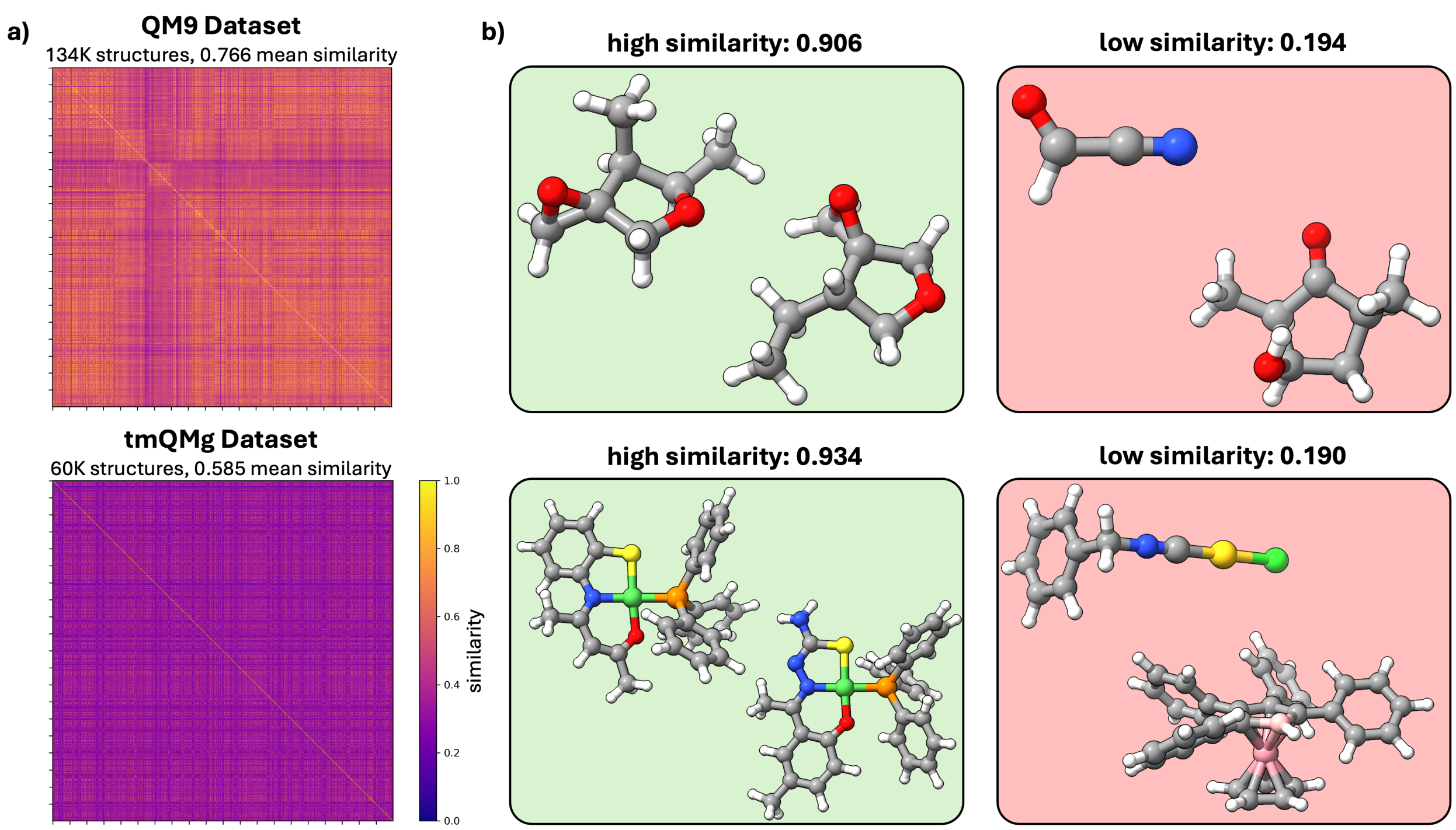

**Figure 3.** Results of similarity and clustering analysis on QM9 (organic chemistry) and tmQMg (inorganic chemistry) datasets. a) Similarity matrices for QM9 (top) and tmQMg (bottom) datasets, where purple indicates low similarity and yellow indicates high similarity in spectral fingerprints. Matrices plotted by randomly down-sampling to 1,000 structures. b) Randomly sampled pairs of structures with high (left) and low (right) spectral similarity. Structure pairs in the top row are extracted from QM9 and include the high-similarity pair (GDB 94813, GDB 116208) and the low-similarity pair (GDB 27, GDB 86142). Structures in the bottom row are from tmQMg include the high-similarity Ni coordination complexes (BEPNOZ, SUXZUI) and the low-similarity Au and Co complexes (BESYOQ, KAGPAJ). Both pairs of high-similarity structures are assigned to the same Leiden cluster; low-similarity structures are assigned to different clusters. Atoms colored as follows: H in white, C in gray, N in blue, O in red, P in orange, S in yellow, Cl in green, Co in pink, Ni in light green, Au in gold.

We observe strong performance on spectral similarity-based clustering of the lipids subset of the QCell dataset, with 96.2% of clusters resulting in distinct, statistically significant distributions of dipole moment magnitudes. In all instances, 100% of the structures considered are successfully parsed by our spectral fingerprints, demonstrating robustness to uncommon atom types or ambiguous connectivity that can cause challenges for standard cheminformatics approaches. Spectral fingerprints serve as a robust baseline that maintains high parsability while identifying chemically meaningful clusters.

Our spectral fingerprints overall demonstrate effective performance on chemically challenging complexes with more diverse atom types and molecular sizes than those observed in organic chemistry. Next, we benchmark the quality of clusters identified by our spectral fingerprints against those from several representative molecular representations (Supporting Information Text S2). We first consider clusters based on the widely-used Bemis–Murcko scaffolds, which encode molecular motifs using a chemical backbone without any latent representation. We also evaluate Morgan fingerprints, which are bit-wise fingerprints encoding the presence or absence of circular substructures in 2D molecular graphs. Revised autocorrelations are continuous fingerprints defined as products and differences of key atomic properties at various depths on a molecular graph, and include both 2D (RACs) and 3D (CD-RACs) variants, comprising the final two fingerprint baselines considered.[15,16]

In terms of statistical significance between clusters, scaffold splits are the worst performing method overall on QM9 and tmQMg, a result to be expected from the reductive encoding of chemistry only in terms of backbones and side chains. Scaffold clusters are further limited to only considering molecules parsable by RDKit. While 84.2% of the QM9 entries considered are parsable as scaffolds, only 70.9% of tmQMg and 0.4% of the QCell set are parsable, highlighting that fingerprints based on 2D molecular graphs may struggle to generalize to inorganic and biological datasets with less canonical connectivity (Table 1). Evaluated on the parsable subset of each dataset, scaffold splits still exhibit performance inferior to most other fingerprints considered (Table 1). This poor performance is partly attributable to the high number of singleton classes identified by scaffold-based clustering.

While the Leiden clustering algorithm applied to spectral fingerprints results in a reasonable number of well-populated clusters, clustering QM9 according to Bemis–Murcko scaffolds yields 1,132 clusters versus 76 from the spectral similarity measure (Table 1). Morgan fingerprints perform well across datasets, reaching performance comparable to our spectral fingerprints and exhibiting the strongest performance of any method on tmQMg. However, they suffer from the same parsability issues plaguing scaffold clusters. While 94.7% of QM9 entries are parsable, only 72.6% of tmQMg and 20.0% of QCell structures are successfully read by RDKit into valid Morgan fingerprints (Table 1). These findings highlight the fact that scaffold- and Morgan fingerprint-based representations are challenging to extend in practice beyond small-molecule organic chemistry to more complex inorganic complexes or biological macromolecules due to limitations on parsability for more diverse structures.

Both variants of revised autocorrelations outperform scaffold splits, with the 3D CD-RACs achieving performance comparable to spectral and Morgan clusters on the QM9 and QCell benchmarks (Table 1). However, poor performance is achieved on tmQMg by CD-RACs, potentially due to CD-RACs weighing all bonded atomic interactions equally and omitting noncovalent interactions, which could be relevant in transition metal complexes. While RACs and CD-RACs both share the high parsability rate exhibited by our spectral fingerprints, their high asymptotic scaling results in extended runtimes limiting application to large datasets. For all datasets considered, spectral fingerprints are generated in less than three minutes (typically only in seconds), whole RACs vectors consistently require over ten times longer to generate (Supporting Information Table S8). Of all 2D and 3D methods considered, spectral fingerprints exhibit the best overall performance due to their generalization across chemical domains, high parsability rates, and efficient computational scaling. We now benchmark the performance of our methods on more complex chemical systems.

As a demonstration of the utility of our spectral fingerprints and clustering algorithms, we again consider the Transition1x[88] dataset of chemical reactions and the QMOF[89,90] dataset of metal-organic frameworks. Spectral, Morgan, RACs, and CD-RACs fingerprints are generated separately for reactant and product molecules and concatenated. This results in a length-128 spectral fingerprint, length-50 RACs fingerprint, length-40 CD-RACs fingerprint, and a length-4,096 Morgan fingerprint for each reaction. Scaffold splits for chemical reactions are defined here by simply concatenating the reactant and product scaffolds. The reactant and product molecules in the Transition1x dataset overlap heavily with QM9, with both datasets being derived from the larger GDB-17 dataset. Despite this chemical similarity, we observe decreased performance of most methods considered when evaluating clusters in terms of Transition1x activation energies relative to QM9 atomization energies (Table 2). This is consistent with the known challenge of estimating chemical kinetics and reaction energies from equilibrium structures, as all fingerprints considered are derived only from reactant and product geometries, while activation energies are calculated using transition state information. Nevertheless, of the fingerprints that yield 100% parsability, spectral fingerprints achieve the strongest performance on the challenging Transition1x benchmark. This is indicative of a reasonable ability to discriminate between reaction energetics from reactant and product geometries alone. Even stronger performance is observed on the QMOF dataset, with 82.7% of cluster pairs identified by spectral fingerprints determined to have statistically distinct band gaps (Table 2). In contrast, RACs and CD-RACs are both unable to generalize to such macromolecular systems, with between 23.3% and 43.4% of cluster-pairs being statistically significant. Scaffolds and Morgan fingerprints methods fare somewhat better, with 50.3 to 61.7% of clusters being statistically distinct, but these RDKit-based methods are unable to parse the entirety of the dataset considered.

**Table 2.** Key metrics for clustering methods and as evaluated on more complex datasets of reaction (Transition1x) and reticular (QMOF) chemistry, as well as large datasets of 2.7 million molecules (GEMS). Extrinsic cluster quality is evaluated with respect to the percentage of cluster-pairs with statistically significant differences in their property distributions as determined by Mann–Whitney U tests.[100] Intrinsic quality is measured using the Davies–Bouldin index (DBI)[102] and Calinski–Harabasz index (CHI).[103] All metrics are evaluated in target property space. Scaffold splits and Morgan fingerprints are generated using RDKit.[95] RACs[15] and CD-RACs[16] are generated using molSimplify.[92,93] Morgan fingerprint generation on GEMS did not complete in 48 hours. Bold indicates the best result for a given metric, and the arrow indicates what direction corresponds to a better value for a metric.

| Dataset (Domain) | Target (Units) | Representation | U-test (↑) | DBI (↓) | CHI (↑) | Parsable (%) | Clusters |
|---|---|---|---|---|---|---|---|
| Transition1x (Reaction) | Activation Energy (kcal/mol) | Spectral | 62.0 | 424 | 15.0 | 100 | 42 |
| | | Scaffold | **72.5** | $1.88\times10^3$ | **21.0** | 44.1 | 75 |
| | | Morgan | 69.2 | **258** | 15.0 | 88.3 | 14 |
| | | RACs | 58.6 | $2.44\times10^3$ | 14.1 | 100 | 21 |
| | | CD-RACs | 61.2 | 322 | 16.9 | 100 | 24 |
| QMOF (Reticular) | Band Gap (eV) | Spectral | **82.7** | 220 | **125** | 100 | 22 |
| | | Scaffold | 50.3 | 246 | 10.9 | 70.7 | 73 |
| | | Morgan | 61.7 | **79.7** | 99.1 | 78.9 | 16 |
| | | RACs | 43.4 | 131 | 29.8 | 100 | 28 |
| | | CD-RACs | 23.3 | 190 | 5.84 | 100 | 48 |
| GEMS (Biological) | Electronic Energy (eV) | Spectral | **99.6** | **528** | $\mathbf{1.97\times10^5}$ | 100 | 759 |
| | | Scaffold | 88.5 | $2.64\times10^3$ | $4.52\times10^3$ | 8.6 | 88 |
| | | Morgan | --- | --- | --- | --- | --- |
| | | RACs | 99.0 | $4.37\times10^3$ | $7.95\times10^3$ | 100 | 4828 |

| | | CD-RACs | 99.0 | $3.07\times10^3$ | $1.91\times10^4$ | 100 | 4122 |
|---|---|---|---|---|---|---|---|

As a final large-scale test case, we consider the GEMS dataset of 2.7M off-equilibrium structures of protein fragments. While spectral fingerprints, RACs, and CD-RACs all exhibit strong performance, spectral fingerprints exhibit the highest percentage of statistically significant clusters (99.6%) with the best DBI score (528.1), and CHI score ($1.97 \times 10^5$, Table 2). Similar to QCell and Transition1x, scaffold fingerprints are only successfully generated for a minority of the dataset, further underscoring their inability to scale to complex, off-equilibrium chemistry. Morgan fingerprints were unable to scale to the large dataset within the 48-hour runtime allowed for all methods. Together with their performance on organic, inorganic, biological, and reaction chemistry datasets, our spectral fingerprints are established to have superior generalizability across chemical domains, scalability to macromolecules and large datasets, and sensitivity to 3D conformers and stereochemistry desired in a molecular representation.

**3d. Integration with Machine Learning and Cheminformatics Workflows**

Having validated the utility of our spectral fingerprints for database screening, we next investigated whether our representations may be applied to cheminformatics and machine learning. While learned representations from expressive graph neural networks (GNNs) generally outperform handcrafted chemical fingerprints in property prediction tasks[27,28], interpretable molecular representations remain valuable. Specifically, the extent to which molecules nearby in representation space exhibit similar properties is represented by the locality. Locality is the geometric property underlying training-free methods such as nearest-neighbor property estimation, similarity-based retrieval, and applicability-domain analysis.[14,54,55,57,160-162] To evaluate the locality of our spectral fingerprints across representative chemical properties and domains, we perform $k$-nearest-neighbor (k-NN) regression[58], predicting each molecule's property as the mean of the properties of its $k$-nearest-neighbors in spectral fingerprint space. We repeat this process with Morgan, RACs, and CD-RACs fingerprints, comparing the performance among each fingerprint evaluated with its native distance (i.e., Tanimoto for the binary Morgan bit vectors, Euclidean for the continuous spectral, RACs, and CD-RACs vectors) across property prediction tasks (i.e., atomization energy, dipole magnitude, HOMO-LUMO gap, and polarizability) spanning organic (QM9) and inorganic (tmQMg) chemistry. This nonparametric analysis depends entirely on the geometry of the representation space, effectively probing the extent to which the neighborhoods identified by each fingerprint carry physical and chemical meaning.

For each task and dataset considered, we calculate the Spearman rank correlation between the $k$-NN-predicted and true property values at neighborhood size $k \in \{1, 5, 25\}$. Across the values considered, $k = 5$ is observed to maximize the mean rank correlation across all tasks and representations, resulting in the best overall performance for all methods (Supporting Information Tables S9–S11). Spectral fingerprints define the most chemically meaningful neighborhoods for size-extensive properties governed by 3D geometry, achieving the highest rank correlation on QM9 atomization energy ($\rho = 0.999$), QM9 polarizability ($\rho = 0.990$), tmQMg polarizability ($\rho = 0.967$), QM9 HOMO-LUMO gap ($\rho = 0.948$), and QM9 dipole magnitude ($\rho = 0.765$, Figure 4). These observations are consistent with the ability of the spectral representation to encode 3D information via pairwise atomic interactions. Across all seven tasks considered, spectral fingerprints also outperform both the RACs and CD-RACs descriptors, which are the most directly comparable due to their being continuous physics-based representations of similar length (Figure 4). Morgan fingerprints surpass spectral fingerprints only on tmQMg dipole magnitude ($\rho = 0.661$) and HOMO-LUMO gap ($\rho = 0.775$), although we note that Morgan fingerprints are only evaluated on the subset of each dataset which are parsable by RDKit (see Section 3c and Supporting Information Tables S12–S14). While Morgan fingerprints encoding the presence of local substructures better capture properties dominated by specific polar functional groups, such as the molecular dipole moment, spectral fingerprints constructed from the eigenvalues of distance-weighted atomic interaction matrices effectively recover the ordering of extensive properties. This nonparametric analysis based purely on distance in representation space confirms that spectral fingerprints embed molecules in a latent space where geometric proximity corresponds to chemical similarity.

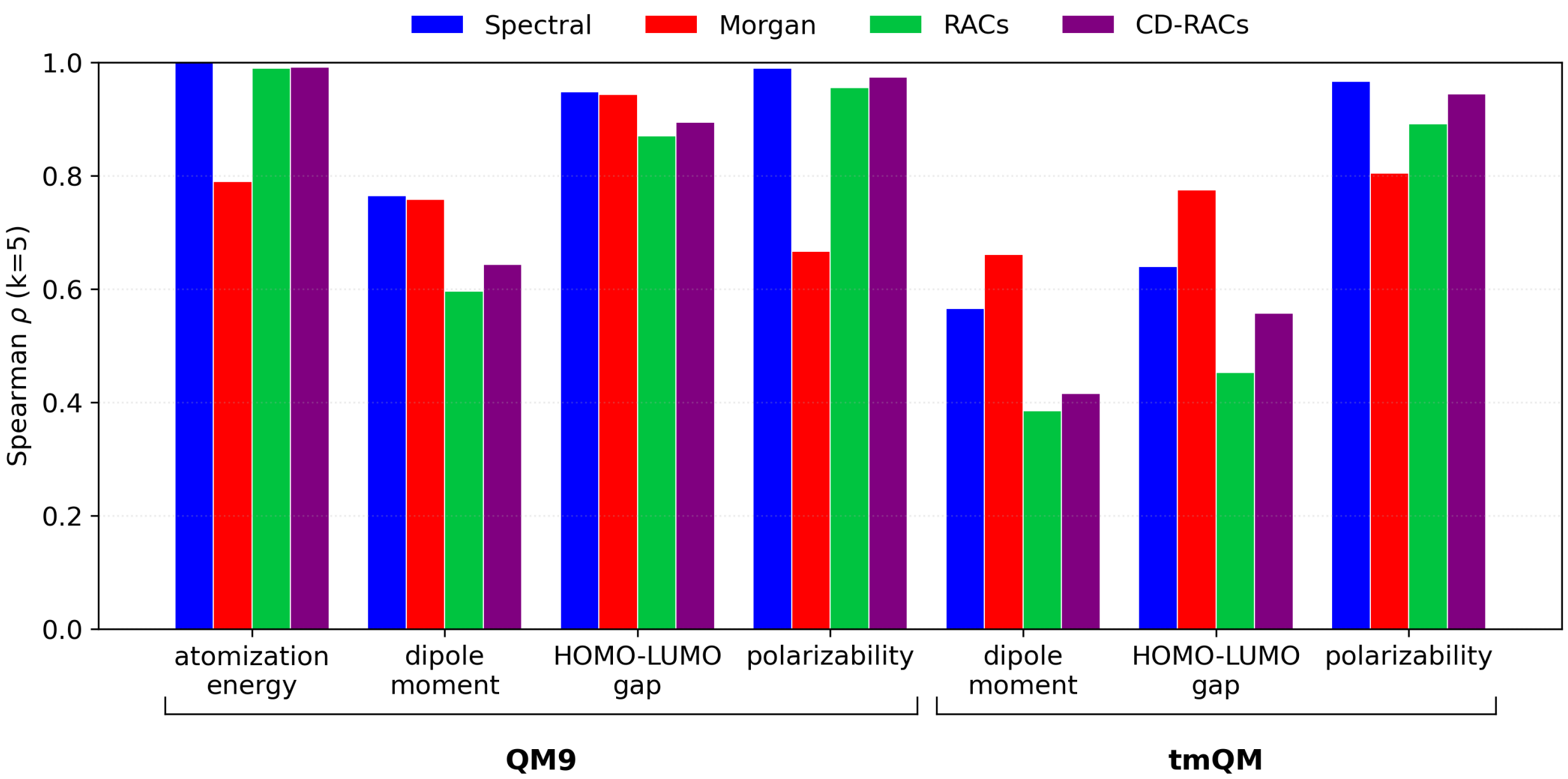

**Figure 4.** Results of $k$-NN regression analysis performed on QM9 and tmQMg datasets in molecular representation space. For a given dataset, target, and representation, Spearman rank correlation is reported between predicted and true property values for each molecule, where predicted values are calculated as the average between the $k = 5$ nearest neighbors in representation space. Spectral, RACs, and CD-RACs are each evaluated on the entire dataset, while Morgan fingerprints are only evaluated on the parsable subset of each dataset (Section 3c).

Beyond *k*-NN-regression, property locality also enables applicability-domain (AD) estimation. AD estimation involves estimating the test set data points on which a trained model is least reliable based on their distance from the training distribution.[59-63,163,164] Towards this end, we train 2D and 3D GNN models on the QM9 and tmQMg datasets (see Sec. 2, Supporting Information Tables S2 and S3). For each trained model, we sort the held-out test set predictions by each molecule's 1-NN distance to the training set in representation space (i.e., spectral, Morgan, RACs, CD-RACs). We analyzed the accumulating error with increasing distance and expect to observe a data coverage-error curve that corresponds to lower errors when only the most similar molecules are predicted. Spectral distance is observed to give the largest low-coverage error reduction among the chemical fingerprints considered in five of the seven tasks considered (Figure 5). Although exact rankings vary when considering subsets of each dataset parsable by RDKit, spectral fingerprints consistently remain competitive (Figure 5). As expected, the model's own latent space distance is the most useful metric, as it has direct access to the learned representation fit to the target data being considered. While the utility of fingerprint distance is more modest for AD estimation than property-locality, our spectral AD estimate is a cheap, model-agnostic structural confidence measure that complements a trained model, as it may be generated prior to training, enabling pre-training insights as to where models are likely to struggle on a given dataset. Together with property locality, spectral fingerprints are established as a practical, training-free component of machine learning workflows that complements existing models.

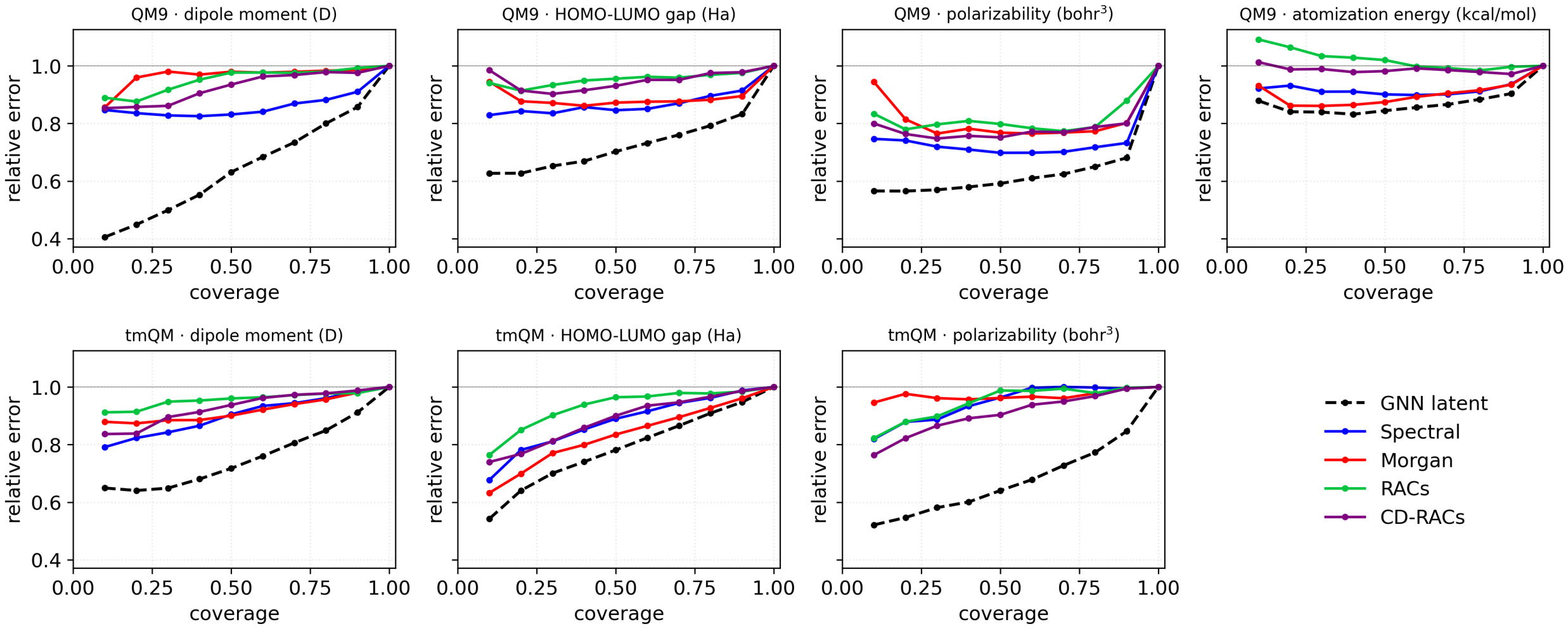

**Figure 5.** Applicability-domain analysis of molecular representations. For each trained model, held-out test molecules are ranked by their distance to the nearest training-set molecule, computed independently in each representation space (i.e., spectral, Morgan, RACs, and CD-RACs). Cumulative RMSE is evaluated over the closest fraction of the test set (i.e., coverage), normalized by the representation's own full test-set (coverage = 1.0) RMSE. The latent space distance of each trained graph neural network (GNN) model is provided for reference.

## 4. Conclusions

In summary, we developed physics-based molecular fingerprints to encode 3D chemical structure using principles from spectral graph theory. To address the gap between fixed-length 2D representations and typically expensive 3D-aware methods, we encoded each molecule as a weighted complete graph over its 3D atomic coordinates, assigning edge weights from four physics-inspired interaction channels (i.e., electrostatics, bonding, sterics, and dispersion). We combined the eigenvalues from the resulting channel-wise graph Laplacians as a fixed-length spectral fingerprint, with similarity defined in terms of the Euclidean distance between fingerprints. Our method is fixed-length, permutation invariant, alignment-free, and interpretable, requiring only 3D atomic coordinates as inputs. We demonstrated that these fingerprints resolve chemically relevant geometric conformers and stereoisomers that are indistinguishable by 2D fingerprints. Strong performance is similarly observed in more challenging systems of chemical reactions and macromolecules (i.e., MOFs and proteins).

We paired our spectral fingerprints with the Leiden community detection algorithm for screening large chemical datasets, benchmarking performance against Bemis–Murcko scaffolds, Morgan fingerprints, RACs, and CD-RACs. Across six datasets spanning organic, inorganic, biological, reaction, and reticular chemistry, spectral fingerprints were consistently observed to exhibit the strongest combination of cluster quality and parsability.

Finally, we evaluated the utility of spectral fingerprints in cheminformatics and machine learning workflows. Nonparametric $k$-nearest-neighbor regression revealed spectral fingerprints to define the most chemically meaningful neighborhoods for size-extensive properties governed by 3D structure. Ranking held-out predictions by spectral distance to the training set likewise produced the largest low-coverage error reduction on most applicability-domain tasks, providing an inexpensive, model-agnostic estimate of prediction reliability prior to model training.

While our spectral fingerprints perform well from small datasets to millions of structures and in regimes from small molecules to macromolecules, future work could investigate algorithms for accelerated handling of large structures. Eigenvalue decomposition of the multichannel Laplacian scales natively as $O(N^3)$ and becomes the computational bottleneck for macromolecules.[165] Because our interaction matrices are symmetric and positive semidefinite by construction and only the extreme eigenvalues are required, Krylov subspace methods such as the Lanczos algorithm may offer numerical advantages to be investigated in future work.[166,167] Finally, while Frobenius normalization enables comparison across diverse and arbitrarily large chemical systems, it introduces degeneracy for molecules with $N = 2$ atoms, though the intended use case of spectral fingerprints is for systems large enough that manual inspection is intractable (i.e., $N > 2$). Further investigations into channel-wise contributions in diverse environments could provide deeper insights regarding chemical interpretability. We anticipate that our spectral fingerprints will serve as a fast, interpretable, and broadly transferable measure of molecular similarity, with applications in dataset curation, chemical space exploration, and pre-processing for machine learning workflows across the chemical sciences.

ASSOCIATED CONTENT

**Supporting Information**. Description of existing chemical datasets used in this work; process used to generate molecular fingerprint baselines; silhouette scores evaluated in target and latent space; hyperparameters used in all machine learning models; performance of all machine learning models; similarity as a function of fingerprint size; similarity as a function of fingerprint size for large chemical systems; channel-wise ablations for representative chemical systems; description of test systems considered; details of *in silico* protein mutations; wall-clock time required to calculate spectral fingerprints; description of metrics used to evaluate cluster quality; cluster evaluation metrics defined in latent space; wall-clock time required to generate all molecular fingerprints; results of $k = 1$ nearest neighbor (NN) regression analysis; results of $k = 5$ NN analysis; results of $k = 25$ NN analysis; results of $k = 1$ NN analysis, controlled for parsability errors; results of $k = 5$ NN analysis, controlled for parsability errors; results of $k = 25$ NN analysis, controlled for parsability errors.

**Data and Software Availability Statement**

An open-source implementation of our spectral fingerprints is available publicly via GitHub at https://github.com/hjkgrp/SpectralScore. All data required to reproduce this work is provided either in the Supporting Information PDF file or in the Zenodo repository at https://doi.org/10.5281/zenodo.21614395.

AUTHOR INFORMATION

**Corresponding Author**

*email:hjkulik@mit.edu

**Notes**

The authors declare no competing financial interest.

ACKNOWLEDGMENT

Funding was provided by a UPI from The Dow Chemical Company. J.W.T. was partially supported by a Leslye Miller Fraser and Darryl M. Fraser Fellowship from the MIT School of Engineering. H.J.K. is supported by a Simon Family Faculty Research Innovation Fund and an Alfred P. Sloan Fellowship in Chemistry. The authors acknowledge the MIT SuperCloud and Lincoln Laboratory Supercomputing Center for providing HPC resources that have contributed to the research results reported in this work. The authors thank Adam H. Steeves for providing a critical reading of the manuscript. The authors thank Teya Bergamaschi and Heecheol Jang for valuable technical discussions and scientific insights.

**For Table of Contents Use Only**

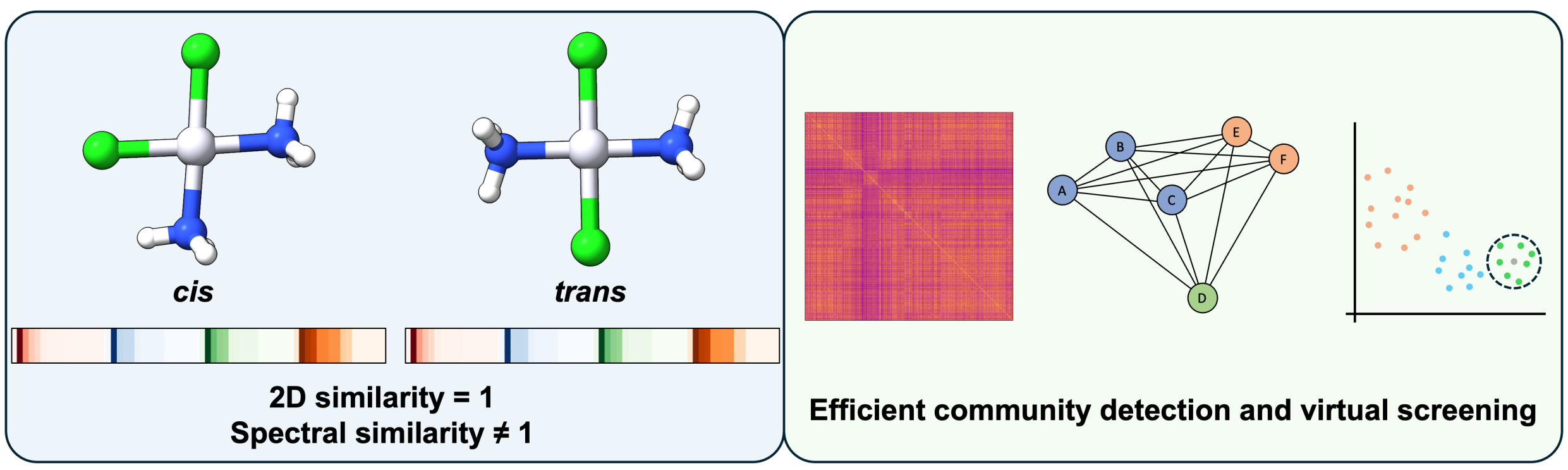

cis
trans
2D similarity = 1
Spectral similarity ≠ 1
A
B
C
D
E
F
Efficient community detection and virtual screening

**Supporting Information for**

***Physics-Based Molecular Fingerprints from Spectral Graph Theory Provide Efficient Geometry-Aware Measures of Chemical Similarity***

Jacob W. Toney[1,2], Ayleen Y. Farnood[1,2], Samir Darouich[1,3,4], and Heather J. Kulik[1,2,5,*]

[1]*Department of Chemical Engineering, Massachusetts Institute of Technology, Cambridge, MA 02139, USA*

[2]*Center for Computational Science and Engineering, Massachusetts Institute of Technology, Cambridge, MA 02139, USA*

[3]*Institute for Theoretical Chemistry, University of Stuttgart, 70569 Stuttgart, Germany*

[4]*Institute for Artificial Intelligence, University of Stuttgart, 70569 Stuttgart, Germany*

[5]*Department of Chemistry, Massachusetts Institute of Technology, Cambridge, MA 02139, USA*

*corresponding author email: hjkulik@mit.edu

*email: hjkulik@mit.edu

**Contents**

**Text S1.** Description of existing chemical datasets used in this work.
QM9 is a collection of 133,885 small organic molecules including up to 9 heavy atoms with only H, C, N, O, and F atom types. Geometries and electronic properties are calculated at the B3LYP/6-31G(2df,p) level of theory.[1] tmQMg includes 60,799 mononuclear transition metal complexes (TMCs) from the Cambridge Structural Database (CSD) with geometries optimized and electronic properties calculated at the PBE-D3BJ/def2-SVP//PBE0-D3BJ/def2-TZVP level of theory.[2] The QCell dataset includes molecular dynamics trajectories of small biological molecules and fragments (e.g., lipids, carbohydrates, nucleic acids, etc.) with electronic properties calculated at the PBE0+MBD(-NL) level of theory.[3] Transition1x includes geometries of reactants, products, and transition states for 10,073 small, neutral, organic reactions calculated at the ωB97x/6–31 G(d) level of theory.[4] The QMOF dataset reports geometries and electronic properties of 20,372 experimentally synthesized metal-organic frameworks calculated with PBE-D3BJ.[5,6] Finally, GEMS includes 2,713,986 off-equilibrium geometries of small protein fragments calculated with PBE0/def2-TZVPP.[7]

**Text S2.** Process used to generate molecular fingerprint baselines.

**Bemis-Murcko Scaffolds and Morgan Fingerprints from RDKit**

Bemis-Murcko scaffolds and Morgan fingerprints were generated as implemented in RDKit version 2026.3.1.[8] For both representations, 3D atomic coordinates were first read using the `Chem.rdmolfiles.MolFromXYZFile` function in RDKit, with connectivity interpreted with `Chem.rdDetermineBonds.DetermineConnectivity` and `Chem.SanitizeMol`. Scaffolds were generated using `Chem.Scaffolds.MurckoScaffold.GetScaffoldForMol` followed by `Chem.MolToSmiles` to return the corresponding SMILES strings. Morgan fingerprints were generated using `Chem.GetMorganGenerator` with a radius of 3, corresponding to the default value in RDKit. Structures which could not be parsed by RDKit or for which molecular representations could not be generated (e.g., failure in `GetScaffoldForMol`) are excluded from the corresponding analysis and reflected in the reported parsability rates (Section 3).

**RACs and CD-RACs from molSimplify**

RACs were generated by computing products and differences in atomic properties (i.e., nuclear charge, Pauling electronegativity, coordination number, covalent radius, and atomic identity) up to depth $d$ and averaging the resulting products and differences across all atoms in the molecule. Specific TMC RACs that apply for octahedral TMCs (i.e., ligand-centered and metal-centered) are not used, even for TMC tasks such as tmQM, which contains non-octahedral structures. We use the default value in molSimplify of $d = 4$. CD-RACs are defined similarly but are scaled with interatomic distance and were calculated here using the molSimplify default value of $d = 3$. Both RACs and CD-RACs were generated using molSimplify version 1.8.0.[9]

**Table S1.** Silhouette scores evaluated in target and latent space. Scaffold splits and Morgan fingerprints are generated using RDKit.[8] RACs[10] and CD-RACs[11] were generated using molSimplify.[9,12] Scaffold splits do not have a latent space representation. Morgan fingerprint generation on GEMS did not complete in 48 hours. Bold indicates the best result for a given metric, and the arrow indicates what direction corresponds to a better value for a metric.

| Dataset (Domain) | Target (Units) | Representation | Silhouette Score, Target Space (↑) | Silhouette Score, Latent Space (↑) | Parsable (%) | Clusters |
|---|---|---|---|---|---|---|
| QM9 (Organic) | Atomization Energy (kcal/mol) | Spectral | -0.39 | **0.12** | 100 | 76 |
| | | Scaffold | -0.65 | --- | 84.2 | 1132 |
| | | Morgan | **-0.24** | 0.01 | 94.7 | 16 |
| | | RACs | -0.35 | 0.01 | 100 | 34 |
| | | CD-RACs | -0.35 | -0.01 | 100 | 34 |
| tmQMg (inorganic) | Gap (eV) | Spectral | **-0.26** | **0.09** | 100 | 31 |
| | | Scaffold | -0.68 | --- | 70.9 | 230 |
| | | Morgan | -0.28 | -0.03 | 72.6 | 22 |
| | | RACs | -0.96 | -0.34 | 100 | 119 |
| | | CD-RACs | -0.82 | -0.09 | 100 | 68 |
| QCell (Biological) | Dipole Magnitude (D) | Spectral | -0.27 | 0.25 | 100 | 13 |
| | | Scaffold | -0.01 | --- | 0.4 | 3 |
| | | Morgan | **0.04** | **0.99** | 20.0 | 3 |
| | | RACs | -0.33 | 0.97 | 100 | 35 |
| | | CD-RACs | -0.33 | 0.97 | 100 | 31 |
| Transition1x (Reaction) | Activation Energy (kcal/mol) | Spectral | -0.25 | **0.20** | 100 | 42 |
| | | Scaffold | -0.65 | --- | 44.1 | 75 |
| | | Morgan | -0.23 | 0.02 | 88.3 | 14 |
| | | RACs | **-0.14** | 0.07 | 100 | 21 |
| | | CD-RACs | -0.20 | 0.05 | 100 | 24 |
| QMOF (Reticular) | Band Gap (eV) | Spectral | **-0.36** | **0.11** | 100 | 22 |
| | | Scaffold | -0.62 | --- | 70.7 | 73 |
| | | Morgan | -0.59 | -0.11 | 78.9 | 16 |
| | | RACs | -0.83 | -0.10 | 100 | 28 |
| | | CD-RACs | -0.87 | -0.15 | 100 | 48 |
| GEMS (Biological) | Electronic Energy (eV) | Spectral | -0.70 | 0.19 | 100 | 759 |
| | | Scaffold | **-0.62** | --- | 8.6 | 88 |
| | | Morgan | --- | --- | --- | --- |
| | | RACs | -0.71 | **0.53** | 100 | 4828 |
| | | CD-RACs | -0.71 | 0.50 | 100 | 4122 |

**Table S2.** Hyperparameters used in all machine learning models trained in this work. Values reflect system defaults in the ElemeNet software package.[13] GNN models utilize a 2D GNN encoder, while EGNN models utilized an E(3)-equivariant 3D encoder. MLP models use a multilayer perceptron readout architecture for the learned representations, while TF models utilize an attention-based transformer architecture.

| hyperparameter | GNN-MLP | GNN-TF | EGNN-MLP | EGNN-TF |
|---|---|---|---|---|
| learning rate | 1e-4 | 1e-4 | 1e-4 | 1e-4 |
| weight decay | 0 | 0 | 0 | 0 |
| encoder layers | 3 | 3 | 3 | 3 |
| encoder neurons | 128 | 128 | 128 | 128 |
| encoder shape | constant | constant | constant | constant |
| encoder dropout | 0.1 | 0.1 | 0.1 | 0.1 |
| encoder activation | relu | relu | relu | relu |
| encoder convolution | graphconv | graphconv | graphconv | graphconv |
| encoder pooling | mean | mean | mean | mean |
| EGNN inverse sublayers | N/A | N/A | 2 | 2 |
| EGNN attention | N/A | N/A | False | False |
| EGNN Gaussians | N/A | N/A | True | True |
| EGNN aggregation | N/A | N/A | 64 | 64 |
| readout layers | 3 | 3 | 3 | 3 |
| readout neurons | 128 | 128 | 128 | 128 |
| readout shape | constant | constant | constant | constant |
| readout dropout | 0.1 | 0.1 | 0.1 | 0.1 |
| readout activation | relu | relu | relu | relu |
| readout norm | True | True | True | True |
| graph attribute hidden size | None | None | None | None |
| transformer attention heads | N/A | 8 | N/A | 8 |
| transformer expansion | N/A | 4 | N/A | 4 |

**Table S3.** Performance of all machine learning models trained in this work. All values are mean absolute error evaluated on the test set.

| Dataset | property (units) | GNN-MLP | GNN-TF | EGNN-MLP | EGNN-TF |
|---|---|---|---|---|---|
| QM9 | atomization energy (kcal/mol) | 7.51 | 1.36 | 2.27 | 0.66 |
| QM9 | dipole magnitude (debye) | 0.491 | 0.459 | 0.141 | 0.163 |
| QM9 | HOMO-LUMO gap (hartree) | $5.54\times10^{-3}$ | $4.44\times10^{-3}$ | $2.64\times10^{-3}$ | $2.45\times10^{-3}$ |
| QM9 | polarizability (bohr$^3$) | 0.496 | 0.270 | 0.152 | 0.117 |
| tmQMg | dipole magnitude (debye) | 1.82 | 1.88 | 1.03 | 1.23 |
| tmQMg | HOMO-LUMO gap (hartree) | 0.0107 | 0.0116 | $9.39\times10^{-3}$ | $9.98\times10^{-3}$ |
| tmQMg | polarizability (bohr$^3$) | 12.73 | 8.81 | 5.44 | 5.34 |

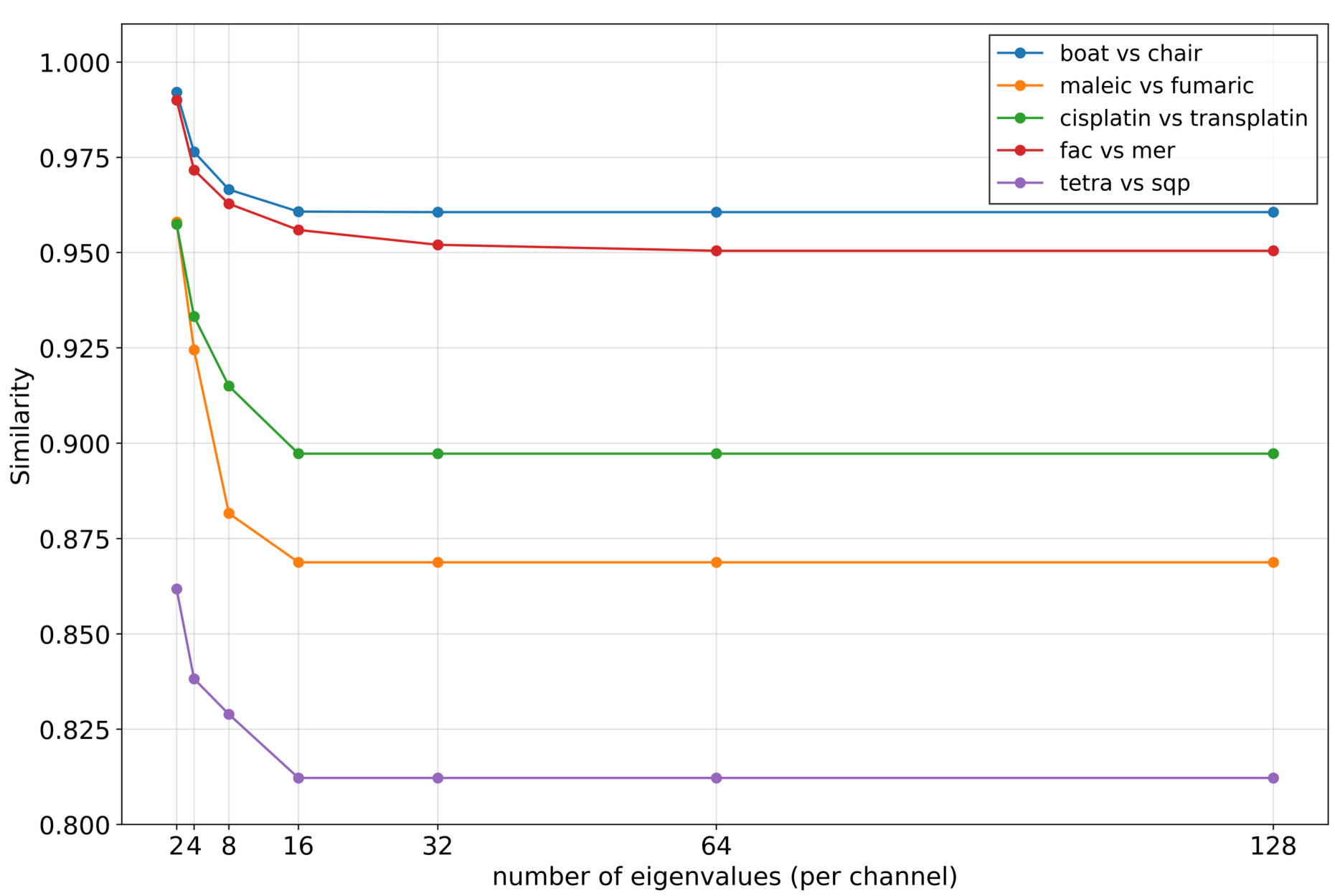

**Figure S1.** Similarity as a function of fingerprint size for representative chemical systems, as indicated in inset legend. All pairs of systems considered have identical graphs but distinct 3D structure. The number of eigenvalues refers to the per-channel length across each of the four channels for electrostatics, dispersion, bonding, and sterics (e.g., selecting 8 eigenvalues results in a length 32 spectral fingerprint). Across systems studied, similarity is observed to plateau at approximately 16 eigenvalues (i.e., a length 64 spectral fingerprint).

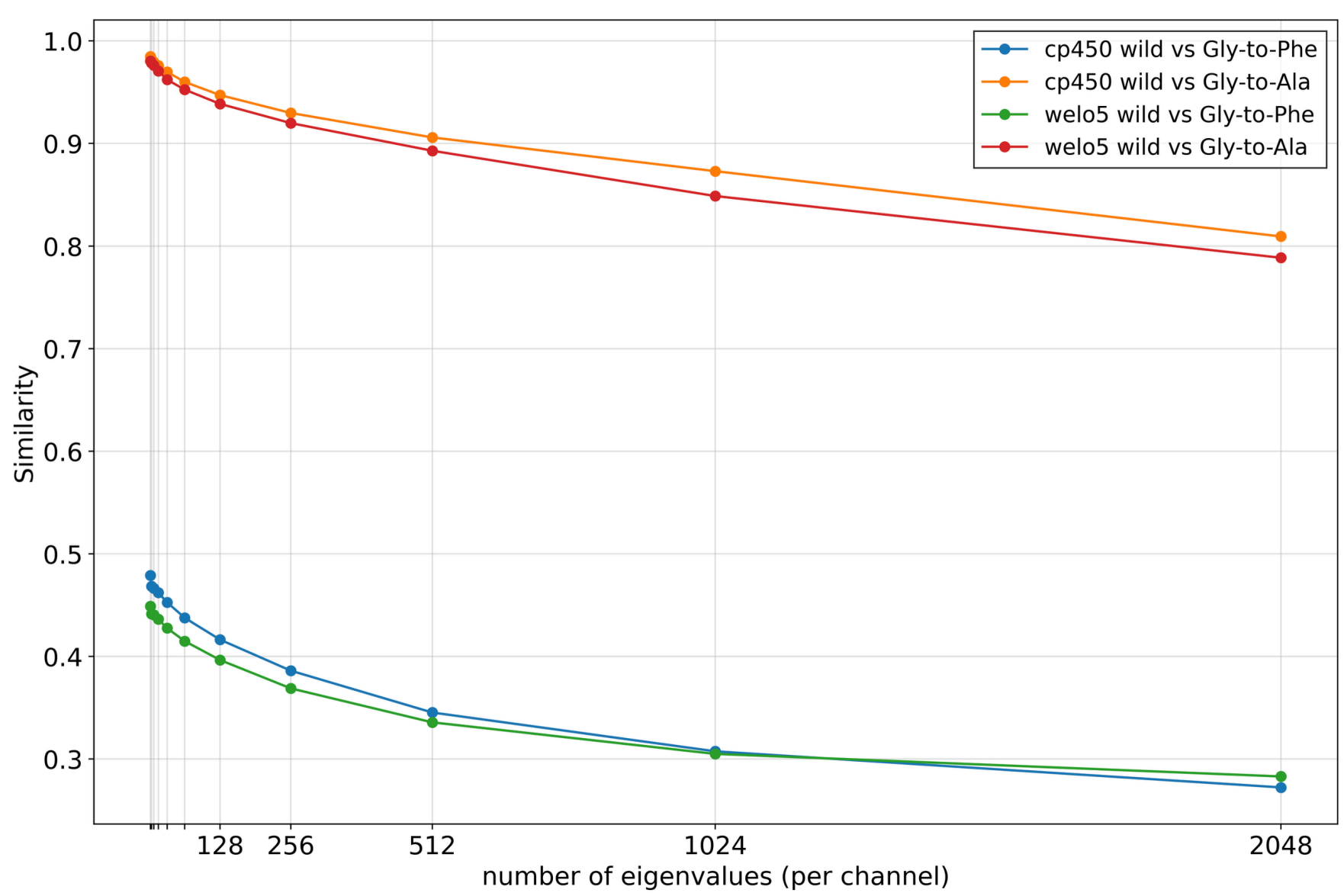

**Figure S2.** Similarity as a function of fingerprint size for large chemical systems, as indicated in inset legend. All pairs of systems are large proteins with *in silico* mutations (Table S4). The number of eigenvalues refers to the per-channel length across each of the four channels for electrostatics, dispersion, bonding, and sterics. Results are observed to be dependent on the number of eigenvalues considered for larger chemical systems.

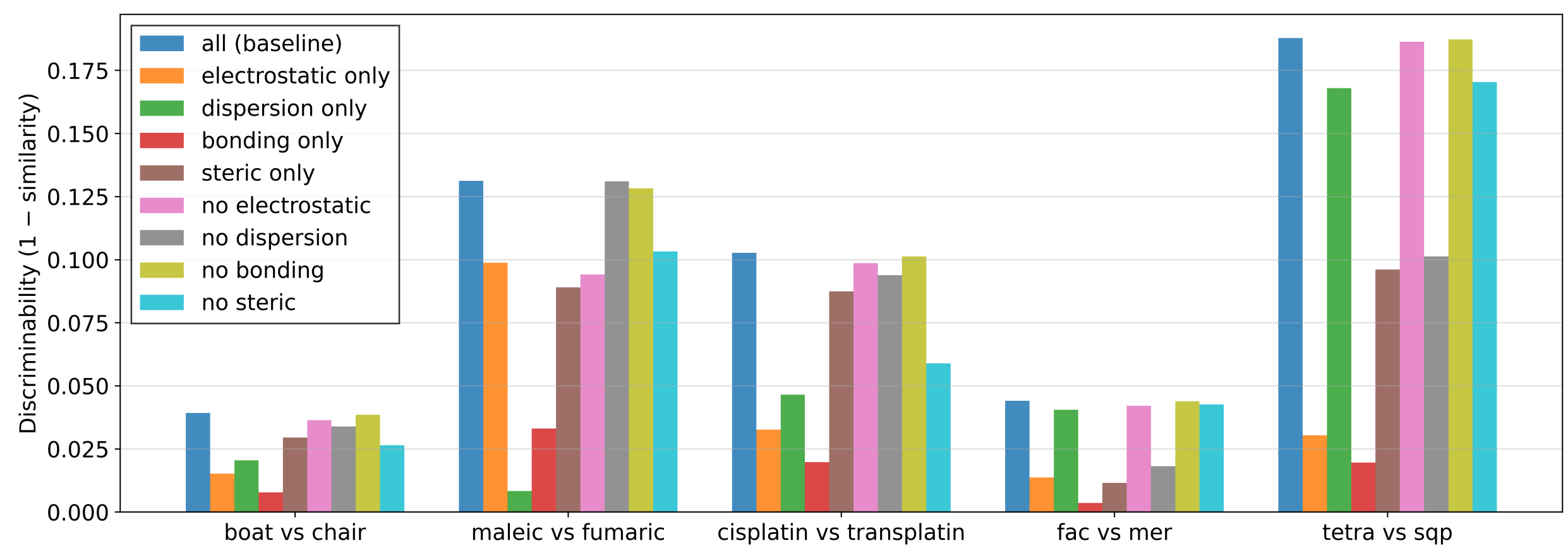

**Figure S3.** Channel-wise ablations for representative chemical systems. Across systems studied, the relative signal provided by each channel varies. All analysis performed with 16 eigenvalues (i.e., a length 64 spectral fingerprint).

**Table S4.** Description of test systems considered in developing and evaluating spectral fingerprints. All structures from the boat-chair, tetra-sqp, cis-trans (organic), cis-trans (inorganic), and fac-mer test cases were prepared manually and optimized with DFT (see Methods). Structures from all remaining test cases were extracted from the reported datasets or databases (i.e., Transition1x[4], QMOF[5], or PDB[14]).

| test case | domain | description |
|---|---|---|
| boat-chair | organic | Boat and chair conformers of cyclohexane. |
| tetra-sqp | inorganic | Tetrahedral and square planar geometries of bis-(propane-1,3-dionato)Ni(II). |
| cis-trans (organic) | organic | Cis and trans isomers of butenedioic acid. |
| cis-trans (inorganic) | inorganic | Cis and trans isomers of $Pt(II)Cl_2(NH_3)_2$. |
| fac-mer | inorganic | Fac and mer isomers of $Ir(III)(ppy)_3$. |
| rxn 3537-3878 | reaction | Reactions 3537 and 3878 in Transition1x dataset. |
| rxn 0728-3537 | reaction | Reactions 0728 and 3537 in Transition1x dataset. |
| rxn 0728-3878 | reaction | Reactions 0728 and 3878 in Transition1x dataset. |
| Zn-Zn MOFs | reticular | Two Zn MOFs (CSD code ACOLIP and HAXMUQ) from QMOF dataset. |
| Gd-Zn1 MOF | reticular | Gd and Zn MOFs (CSD code HAVFAN and ACOLIP) from QMOF dataset. |
| Gd-Zn2 MOF | reticular | Gd and Zn MOFs (CSD code HAVFAN and HAXMUQ) from QMOF dataset. |
| P450, small change | biological | Wild-type and Gly-to-Ala mutant of cytochrome P450 protein (PDB ID 8U1I). |
| P450, large change | biological | Wild-type and Gly-to-Phe mutant of cytochrome P450 protein (PDB ID 8U1I). |
| WelO5, small change | biological | Wild-type and Gly-to-Ala mutant of WelO5 protein (PDB ID 5J4R). |
| WelO5, large change | biological | Wild-type and Gly-to-Phe mutant of WelO5 protein (PDB ID 5J4R). |

**Table S5.** Details of *in silico* protein mutations. All mutations performed on Cytochrome P450 (PDB ID: 8U1I) and WelO5 protein (PDB ID: 5J4R). The number of glycine (Gly), alanine (Ala), and phenylalanine (Phe) residues for the wild-type and mutant proteins are reported. "Gly-to-Ala mutant" has all glycine residues replaced with alanine, while "Gly-to-Phe mutant" has all glycine residues replaced with phenylalanine. All substitutions were performed with ChimeraX v1.6.1.[15]

| protein | name | number of atoms | GLY count | ALA count | PHE count |
|---|---|---|---|---|---|
| Cytochrome P450 | wild-type | 3,653 | 31 | 45 | 15 |
| | Gly-to-Ala mutant | 3,684 | 0 | 76 | 15 |
| | Gly-to-Phe mutant | 3,870 | 0 | 45 | 46 |
| WelO5 | wild-type | 2,439 | 23 | 19 | 15 |
| | Gly-to-Ala mutant | 2,462 | 0 | 42 | 15 |
| | Gly-to-Phe mutant | 2,600 | 0 | 19 | 38 |

**Table S6.** Wall-clock time required to calculate spectral fingerprints and compute pairwise similarity scores across test cases with various numbers of atoms. All operations performed on an M1 MacBook Pro with 16 GB of RAM.

| test case | total atoms | similarity | wall-clock time (ms) |
|---|---|---|---|
| boat-chair | 36 | 0.9608 | 2.05 |
| tetra-sqp | 34 | 0.8122 | 12.6 |
| cis-trans (organic) | 24 | 0.8688 | 8.62 |
| cis-trans (inorganic) | 22 | 0.8973 | 6.23 |
| fac-mer | 61 | 0.9559 | 16.3 |
| rxn 3537-3878 | 32 | 0.7031 | 9.03 |
| rxn 0728-3537 | 38 | 0.4338 | 3.57 |
| rxn 0728-3878 | 38 | 0.4290 | 1.46 |
| Zn-Zn MOFs | 192 | 0.7401 | 29.3 |
| Gd-Zn1 MOF | 212 | 0.2674 | 7.17 |
| Gd-Zn2 MOF | 202 | 0.2774 | 6.04 |
| P450, small change | 7,337 | 0.9759 | 3.06e4 |
| P450, large change | 7,523 | 0.4621 | 3.24e4 |
| WelO5, small change | 4901 | 0.9706 | 1.15e4 |
| WelO5, large change | 5039 | 0.4363 | 1.20e4 |

**Text S3.** Description of metrics used to evaluate cluster quality and fingerprint generality.

**Mann–Whitney U test**

A Mann–Whitney U test is a non-parametric statistical significance test used to compare two independent distributions without assuming distributions are normally distributed. The null hypothesis that the two sets of samples come from the same distribution is rejected or not rejected based on the test statistic and chosen significance level. In all experiments, we report the percentage of clusters determined to be statistically significant (i.e., we consider the clusters distinct and reject the null hypothesis) at the $\alpha = 0.05$ significance level.

**Davies–Bouldin index**

The Davies–Bouldin index (DBI) is used to evaluate cluster quality. It is defined as the average ratio of scatter within a cluster (i.e., dispersion) to separation between different clusters. The DBI is strictly non-negative, with lower values representing better quality clusters.

**Calinski–Harabasz index**

The Calinski–Harabasz index (CHI) is another cluster evaluation metric, defined in terms of the inter-cluster variance to intra-cluster variance. Larger values indicate better cluster quality, and the clusters are both distinct from one another and compact.

**Parsability**

Parsability refers to the percentage of molecules for which fingerprints were successfully generated for a given method.

**Table S7.** Cluster evaluation metrics defined in latent space. Intrinsic quality is measured using the Davies-Bouldin index (DBI)[16] and Calinski-Harabasz index (CHI).[17] Morgan fingerprints are generated using RDKit.[8] RACs[10] and CD-RACs[11] are generated using molSimplify.[9,12] Scaffold splits do not have a latent space representation and are therefore excluded. Bold indicates the best result for a given metric, and the arrow indicates what direction corresponds to a better value for a metric.

| Dataset (Domain) | Target (Units) | Representation | DBI (↓) | CHI (↑) | Parsable (%) | Clusters |
|---|---|---|---|---|---|---|
| QM9 (Organic) | Atomization Energy (kcal/mol) | Spectral | 1.96 | $6.54\times10^{3}$ | 100 | 76 |
| | | Morgan | 5.88 | 927 | 94.7 | 16 |
| | | RACs | **1.83** | $1.25\times10^{4}$ | 100 | 34 |
| | | CD-RACs | 2.11 | $\mathbf{1.62\times10^{4}}$ | 100 | 34 |
| tmQMg (Inorganic) | Gap (eV) | Spectral | 2.11 | $4.88\times10^{3}$ | 100 | 31 |
| | | Morgan | 6.70 | 133 | 72.6 | 22 |
| | | RACs | **1.22** | $2.21\times10^{3}$ | 100 | 119 |
| | | CD-RACs | 1.29 | $\mathbf{5.58\times10^{3}}$ | 100 | 68 |
| QCell (Biological) | Dipole Magnitude (D) | Spectral | 1.60 | $4.04\times10^{4}$ | 100 | 13 |
| | | Morgan | 0.568 | $8.74\times10^{3}$ | 20.0 | 3 |
| | | RACs | 0.280 | $4.50\times10^{7}$ | 100 | 35 |
| | | CD-RACs | **0.139** | $\mathbf{9.65\times10^{7}}$ | 100 | 31 |
| Transition1x (Reaction) | Activation Energy (kcal/mol) | Spectral | 1.57 | $1.07\times10^{3}$ | 100 | 42 |
| | | Morgan | 4.93 | 97.0 | 88.3 | 14 |
| | | RACs | **1.53** | $1.30\times10^{3}$ | 100 | 21 |
| | | CD-RACs | 1.66 | $\mathbf{1.36\times10^{3}}$ | 100 | 24 |
| QMOF (Reticular) | Band Gap (eV) | Spectral | 1.85 | $1.08\times10^{3}$ | 100 | 22 |
| | | Morgan | 6.47 | 35.1 | 78.9 | 16 |
| | | RACs | 1.28 | $1.92\times10^{3}$ | 100 | 28 |
| | | CD-RACs | **1.05** | $\mathbf{2.02\times10^{3}}$ | 100 | 48 |
| GEMS (Biological) | Electronic Energy (eV) | Spectral | 1.68 | $\mathbf{9.25\times10^{4}}$ | 100 | 759 |
| | | Morgan | --- | --- | --- | --- |
| | | RACs | **1.09** | $2.53\times10^{4}$ | 100 | 4828 |
| | | CD-RACs | 1.12 | $3.66\times10^{4}$ | 100 | 4122 |

**Table S8.** Wall-clock time required to generate all molecular fingerprints for large chemical databases. All operations for QM9, tmQMg, QCell, Transition1x, and QMOF datasets performed on an M1 MacBook Pro with 16 GB of RAM. Fingerprints for the GEMS dataset were generated on an HPC cluster using 16 CPUs and 256 GB of RAM. Morgan fingerprints were not successfully generated on the GEMS dataset within the 48-hour wall time limit.

| **Dataset (Domain)** | **Spectral** | **Scaffold** | **Morgan** | **RACs** | **CD-RACs** |
|---|---|---|---|---|---|
| QM9 (Organic) | 13 sec | 29 sec | 30 sec | 103 sec | 80 sec |
| tmQMg (Inorganic) | 11 sec | 65 sec | 59 sec | 286 sec | 244 sec |
| QCell (Biological) | 30 sec | 307 sec | 40 sec | 1335 sec | 1239 sec |
| Transition1x (Reaction) | 1 sec | 2 sec | 3 sec | 7 sec | 4 sec |
| QMOF (Reticular) | 11 sec | 57 sec | 17 sec | 394 sec | 346 sec |
| GEMS (Biological) | 185 sec | 263 sec | --- | 985 sec | 675 sec |

**Table S9.** Results of $k = 1$ nearest neighbor (NN) regression analysis. For each molecule, Spearman rank correlation is reported between predicted and true values, where predicted values correspond to those from the $k = 1$ nearest neighbor in representation space. Spectral, RACs, and CD-RACs are each evaluated on the entire dataset, while Morgan fingerprints are only evaluated on the parsable subset of each dataset (Section 3c).

| Dataset (Domain) | Target (Units) | Spectral | Morgan | RACs | CD-RACs |
|---|---|---|---|---|---|
| QM9 (organic) | Atomization Energy (kcal/mol) | **0.999** | 0.598 | 0.988 | 0.922 |
| | Dipole Magnitude (debye) | **0.673** | 0.542 | 0.526 | 0.573 |
| | Gap (hartree) | **0.936** | 0.934 | 0.854 | 0.875 |
| | Polarizability ($bohr^3$) | **0.988** | 0.416 | 0.956 | 0.974 |
| tmQMg (inorganic) | Dipole Magnitude (debye) | 0.530 | **0.625** | 0.330 | 0.403 |
| | Gap (hartree) | 0.600 | **0.741** | 0.393 | 0.508 |
| | Polarizability ($bohr^3$) | **0.956** | 0.761 | 0.861 | 0.927 |

**Table S10.** Results of $k = 5$ nearest neighbor regression analysis. For each molecule, Spearman rank correlation is reported between predicted and true values, where predicted values are calculated as the average between the $k = 5$ nearest neighbors in representation space. Spectral, RACs, and CD-RACs are each evaluated on the entire dataset, while Morgan fingerprints are only evaluated on the parsable subset of each dataset (Section 3c).

| Dataset (Domain) | Target (Units) | Spectral | Morgan | RACs | CD-RACs |
|---|---|---|---|---|---|
| QM9 (organic) | Atomization Energy (kcal/mol) | **0.999** | 0.790 | 0.990 | 0.992 |
| | Dipole Magnitude (debye) | **0.765** | 0.758 | 0.597 | 0.643 |
| | Gap (hartree) | **0.948** | 0.943 | 0.871 | 0.895 |
| | Polarizability ($bohr^3$) | **0.990** | 0.667 | 0.956 | 0.974 |
| tmQMg (inorganic) | Dipole Magnitude (debye) | 0.566 | **0.661** | 0.385 | 0.416 |
| | Gap (hartree) | 0.640 | **0.775** | 0.453 | 0.558 |
| | Polarizability ($bohr^3$) | **0.967** | 0.804 | 0.891 | 0.945 |

**Table S11.** Results of $k = 25$ nearest neighbor regression analysis. For each molecule, Spearman rank correlation is reported between predicted and true values, where predicted values are calculated as the average between the $k = 25$ nearest neighbors in representation space. Spectral, RACs, and CD-RACs are each evaluated on the entire dataset, while Morgan fingerprints are only evaluated on the parsable subset of each dataset (Section 3c).

| Dataset (Domain) | Target (Units) | Spectral | Morgan | RACs | CD-RACs |
|---|---|---|---|---|---|
| QM9 (organic) | Atomization Energy (kcal/mol) | **0.997** | 0.850 | 0.985 | 0.983 |
| | Dipole Magnitude (debye) | 0.755 | **0.802** | 0.583 | 0.614 |
| | Gap (hartree) | 0.926 | **0.940** | 0.839 | 0.877 |
| | Polarizability ($bohr^3$) | **0.985** | 0.766 | 0.933 | 0.960 |
| tmQMg (inorganic) | Dipole Magnitude (debye) | 0.528 | **0.617** | 0.381 | 0.401 |
| | Gap (hartree) | 0.608 | **0.740** | 0.449 | 0.564 |
| | Polarizability ($bohr^3$) | **0.963** | 0.796 | 0.882 | 0.941 |

**Table S12.** Results of $k = 1$ nearest neighbor regression analysis, controlled for parsability errors. For each molecule, Spearman rank correlation is reported between predicted and true values, where predicted values correspond to those from the $k = 1$ nearest neighbor in representation space. All methods are evaluated on the subset of each dataset parsable by RDKit to generate Morgan fingerprints.

| Dataset (Domain) | Target (Units) | Spectral | Morgan | RACs | CD-RACs |
|---|---|---|---|---|---|
| QM9 (organic) | Atomization Energy (kcal/mol) | **0.999** | 0.598 | 0.988 | 0.993 |
| | Dipole Magnitude (debye) | **0.670** | 0.542 | 0.519 | 0.568 |
| | Gap (hartree) | **0.934** | 0.934 | 0.850 | 0.872 |
| | Polarizability ($bohr^3$) | **0.989** | 0.416 | 0.955 | 0.974 |
| tmQMg (inorganic) | Dipole Magnitude (debye) | 0.535 | **0.625** | 0.356 | 0.392 |
| | Gap (hartree) | 0.584 | **0.741** | 0.379 | 0.486 |
| | Polarizability ($bohr^3$) | **0.954** | 0.761 | 0.858 | 0.927 |

**Table S13.** Results of $k = 5$ nearest neighbor regression analysis, controlled for parsability errors. For each molecule, Spearman rank correlation is reported between predicted and true values, where predicted values are calculated as the average between the $k = 5$ nearest neighbors in representation space. All methods are evaluated on the subset of each dataset parsable by RDKit to generate Morgan fingerprints.

| Dataset (Domain) | Target (Units) | Spectral | Morgan | RACs | CD-RACs |
|---|---|---|---|---|---|
| QM9 (organic) | Atomization Energy (kcal/mol) | **0.999** | 0.790 | 0.990 | 0.992 |
| | Dipole Magnitude (debye) | **0.762** | 0.758 | 0.588 | 0.637 |
| | Gap (hartree) | **0.947** | 0.943 | 0.868 | 0.893 |
| | Polarizability ($bohr^3$) | **0.990** | 0.667 | 0.955 | 0.974 |
| tmQMg (inorganic) | Dipole Magnitude (debye) | 0.575 | **0.661** | 0.422 | 0.434 |
| | Gap (hartree) | 0.630 | **0.775** | 0.445 | 0.546 |
| | Polarizability ($bohr^3$) | **0.967** | 0.804 | 0.889 | 0.945 |

**Table S14.** Results of $k = 25$ nearest neighbor regression analysis, controlled for parsability errors. For each molecule, Spearman rank correlation is reported between predicted and true values, where predicted values are calculated as the average between the $k = 25$ nearest neighbors in representation space. All methods are evaluated on the subset of each dataset parsable by RDKit to generate Morgan fingerprints.

| Dataset (Domain) | Target (Units) | Spectral | Morgan | RACs | CD-RACs |
|---|---|---|---|---|---|
| QM9 (organic) | Atomization Energy (kcal/mol) | **0.997** | 0.850 | 0.985 | 0.983 |
| | Dipole Magnitude (debye) | 0.751 | **0.802** | 0.575 | 0.606 |
| | Gap (hartree) | 0.925 | **0.940** | 0.835 | 0.874 |
| | Polarizability ($bohr^3$) | **0.985** | 0.766 | 0.931 | 0.959 |
| tmQMg (inorganic) | Dipole Magnitude (debye) | 0.545 | **0.617** | 0.429 | 0.437 |
| | Gap (hartree) | 0.602 | **0.740** | 0.439 | 0.557 |
| | Polarizability ($bohr^3$) | **0.963** | 0.796 | 0.879 | 0.942 |